\documentclass[fleqn,usenatbib]{mnras}
\usepackage[switch]{lineno}
\linenumbers
\usepackage{newtxtext,newtxmath}
\usepackage{booktabs}

\usepackage[T1]{fontenc}
\usepackage{color}
\usepackage{xcolor}
\usepackage{hyperref}
\usepackage{orcidlink}

\DeclareRobustCommand{\VAN}[3]{#2}
\let\VANthebibliography\thebibliography
\def\thebibliography{\DeclareRobustCommand{\VAN}[3]{##3}\VANthebibliography}

\usepackage{amsmath}	
\usepackage{graphicx}
\usepackage{epstopdf}  
\usepackage{booktabs}
\usepackage{tabularx}  
\usepackage{pdflscape}
\usepackage{gensymb}
\usepackage[UKenglish]{babel}
\usepackage{fix-cm}

\title[Radio Galaxies detection and characterization using deep learning]{Radio Galaxies detection and characterization using deep learning techniques}

\author[Sanjay Khatik et al.]{
Sanjay Khatik$^{\orcidlink{0009-0000-1182-6252}}$,$^{1}$\thanks{E-mail: sanjaykh@iitk.ac.in (SK), khatiksanjay757@gmail.com}
Rohit Sharma$^{\orcidlink{0000-0003-0485-7098}}$,$^{2}$
and Pankaj Jain$^{\orcidlink{0000-0001-8181-5639}}$ $^{2}$
\\
$^{1}$Department of Physics, Indian Institute of Technology, Kanpur, UP, 208016, India\\
$^{2}$Department of Space, Planetary, Astronomical Sciences \& Engineering,  Indian Institute of Technology, Kanpur, UP, 208016, India
}

\date{Accepted XXX. Received YYY; in original form ZZZ}

\pubyear{\the\year{}}

\begin{document}
\nolinenumbers
\label{firstpage}
\pagerange{\pageref{firstpage}--\pageref{lastpage}}
\maketitle

\begin{abstract}
Future radio telescopes will generate data volumes that are increasingly difficult to analyse using traditional statistical methods, motivating the adoption of machine-learning techniques. In this work, we present \texttt{YOLO-Chars} (\textit{YOLO-based Detection and Characterisation of Radio Sources}), a two-stage deep-learning framework for the automated detection and characterisation of radio galaxies in survey images. The framework is developed and evaluated using the Square Kilometre Array Science Data Challenge 1 (SKA SDC1) dataset. In the first stage, customised YOLO-based multi-scale detection models are used to localise compact and extended sources across large sky maps. In the second stage, a dedicated source-characterisation network estimates the physical properties of the detected sources. We focus on three key parameters: flux density, angular size, and position angle. Our results show that \texttt{YOLO-Chars} achieves competitive detection and characterisation performance on the SKA SDC1 benchmark, demonstrating its potential as a scalable framework for next-generation radio continuum surveys.
\end{abstract}



\section{Introduction} \label{Introduction}
The next generation of radio telescopes, such as the Square Kilometre Array (SKA), will probe the universe with unprecedented data quality \citep{2026arXiv260620366B}. SKA will generate enormous volumes of radio data. Traditional methods for detecting and characterising radio sources may not be efficient or scalable enough to process such massive volumes of data. This has created a growing need for techniques like Machine Learning (ML) and Deep Learning (DL), which are well-suited for automatically analysing large-scale astronomical data.

DL has transformed computer vision by enabling models to learn hierarchical representations from raw images through convolutional neural networks (CNNs; LeCun et al. 2015). Landmark CNNs such as \texttt{AlexNet} \citep{Krizhevsky2012}, \texttt{VGG} \citep{Simonyan_Zisserman2015}, and \texttt{ResNet} \citep{he_2015_ResNet} excelled in image classification, paving the way for object detection—another core task that localises and classifies multiple objects by predicting bounding boxes and labels.
Object detection methods are broadly divided into two categories. Region-based (two-stage) detectors, such as \texttt{R-CNN} \citep{Girshick_rcnn2014}, \texttt{Fast R-CNN} \citep{Girshick_FastRCnn2015}, and \texttt{Faster R-CNN} \citep{Ren_FasterRCnn2015}, first generate region proposals and then refine them in a second stage, offering high accuracy, but at the cost of significant computational overhead from repeated processing. In contrast, single-stage detectors predict classes and bounding boxes directly in one forward pass by dividing the image into a fixed grid and treating detection as a regression problem over grid cells. Key examples include \texttt{SSD} \citep{Liu2016} and the \texttt{YOLO} family \citep{Redmon_YOLOv1_2016, Redmon_YOLOv2_2017, RedmonFarhadi_YOLOv3_2018}. These single-stage approaches are especially valuable in radio astronomy, where rapid inference is essential for processing massive image volumes and diverse source morphologies (compact to extended) from next-generation continuum surveys such as ASKAP EMU, MeerKAT, and SKA precursors.

In recent years, DL techniques have been increasingly applied in radio astronomy to automate source detection, morphological classification, and characterisation, addressing challenges posed by faint, extended, complex, and overlapping sources in large continuum surveys. Early CNN-based methods demonstrated the potential of deep learning for these tasks. \texttt{DeepSource} \citep{VafaeiSadr2019} detected point sources in simulated MeerKAT images. \texttt{ConvoSource} \citep{Lukic2019} reconstructed image cutouts to locate sources. \texttt{CLARAN} \citep{Wu_claran2019} classified complex morphologies in Radio Galaxy Zoo data using an enhanced Faster R-CNN with multi-wavelength support, although its two-stage design limited scalability for large surveys. \citet{Aniyan_2017} applied CNNs to classify Fanaroff-Riley and bent-tailed radio galaxies from VLA FIRST survey images. \citet{Lukic_2018} used a simple CNN to classify compact versus extended sources in Radio Galaxy Zoo data, while \citet{Lukic2019} showed that standard CNNs outperformed capsule networks for morphological classification in LOFAR images due to better noise robustness. Later methods expanded capabilities: \citet{Tang_2022} proposed multi-branch CNNs for FR-I / II classification using NVSS and FIRST data.  \citet{Rustige_2023} curated a FIRST dataset and benchmarked CNNs for the FR-I / II, compact, and bent classes. Lao et al. (2023) introduced HeTu-v2, combining Mask R-CNN and Transformer components for simultaneous segmentation and classification of radio sources. \texttt{JLRAT2} \citep{Yu_Lei_2022} integrated denoising and Feature Pyramid Networks for SDC1 detection/classification, although parameter estimation remained classical. Recent advances incorporate multimodal, weakly-supervised, self-supervised, and vision-language models. RadioGalaxyNET \citep{Gupta2024} provides annotated radio+IR benchmarks for extended galaxy detection and host identification. RG-CAT \citep{rg_cat_2024} used Gal-DINO to unify detection, morphology classification, and infrared host association in ASKAP EMU Pilot data. Weakly-supervised methods \citep{Gupta_2023} reduce pixel-level labelling costs for extended morphologies using class-level labels and refined class activation maps. Self-supervised contrastive learning on EMU/MeerKAT data enables robust detection, classification, and discovery of peculiar objects with minimal annotations \citep{2019PASA...36...37R}. Vision-language models such as radio-llava assist in source description, classification, and anomaly highlighting \citep{riggi2025radiollava}. Unsupervised self-organising maps on ASKAP EMU Pilot, DINGO, and SWAG-X data uncovered rare morphologies, including new candidate Odd Radio Circles \citep{Gupta_Huynh_2022}. Single-stage CNN detectors remain efficient and robust across data regimes. \texttt{YOLO-CIANNA} \citep{Cornu2024} customised a YOLO architecture for detection and characterisation on SDC1 maps, while its extension to 3D hyperspectral HI cubes in SDC2 demonstrates scalability to complex radio data \citep{cornu2026}.

This paper introduces \texttt{YOLO-Chars}, a decoupled DL framework for automatic detection and characterisation of radio sources in crowded SKA SDC1 fields (560 MHz, 1000-hour integration). The proposed framework adopts a practical and modular design in which source detection and source characterisation are independently optimised, providing flexibility and extensibility for future developments. The SDC1 dataset has been previously analysed using traditional source finders (e.g., \texttt{SExtractor} \citep{Bertin_1996}, \texttt{PyBDSF} \citep{Mohan_2015}, \texttt{SELAVY} \citep{Whiting_Humphreys_2012}, \texttt{AEGEAN-2.0} \citep{Hancock_Paul_2018}, \texttt{ProFound} \citep{Robotham_2018}) and DL methods (e.g., \texttt{YOLO-CIANNA} \citep{Cornu2024}, \texttt{CLARAN} \citep{Wu_claran2019}, \texttt{ConvoSource} \citep{Lukic2019}, \texttt{JLRAT2} \citep{Yu_Lei_2022}). \texttt{YOLO-Chars} comprises two custom, sequentially coupled modules: (1) a single-stage detector inspired by YOLOv3 \citep{RedmonFarhadi_YOLOv3_2018} that localises sources by predicting bounding boxes across multiple scales, and (2) a separate lightweight CNN that takes variable-sized source regions defined by the corresponding bounding boxes as input and directly regresses the key source parameters, namely the integrated flux density ($f$), major axis ($b_{\rm maj}$), minor axis ($b_{\rm min}$), and position angle ($\phi$), represented by $\sin(\phi)$ and $\cos(\phi)$. Unlike most prior approaches, which separate detection/classification from parameter estimation and rely on classical fitting, \texttt{YOLO-Chars} employs a dedicated DL-based source characterisation network to directly estimate these source parameters, providing improved integration and flexibility for next-generation radio continuum surveys. For comparison, Appendix~\ref{appendix_3} presents a benchmark between \texttt{YOLO-Chars} and PyBDSF on a common subset of the SKA SDC1 field.

The rest of this paper is organised as follows. Section~\ref{Data_Description} provides a brief overview of the dataset. In Section~\ref{Scoring_procedure}, we describe the evaluation procedure defined for the SKA SDC1 challenge. Section~\ref{Data_Preprocessing} outlines the data preprocessing steps used for both the detection and the characterisation tasks. Section~\ref{detectors} details the network architecture and methodology of the detection models. Section~\ref{Multi_Scale_Characterization} presents the architecture and methodology of the characterisation model. Section~\ref{Loss_function_Setup} describes the loss functions used for optimisation, along with the corresponding training and validation curves. Section~\ref{Analysis_and_Results} reports the main results for both detection and characterisation in their optimal configurations. Finally, Section~\ref{sec:discussion} summarises the key findings, discusses limitations, and provides concluding remarks.

\section{Data Description} \label{Data_Description}
The SKA SDC1 dataset\footnote{\url{https://www.skao.int/en/464/ska-science-data-challenge-1}} dataset consists of nine simulated continuum images in FITS format, covering three frequency bands:
\begin{itemize}
    \item  SKA Mid Band 1 -- \textbf{560 MHz}
    \item  SKA Mid Band 2 -- \textbf{1.4 GHz}
    \item  SKA Mid Band 3 -- \textbf{9.2 GHz}
\end{itemize}
Each frequency includes simulations generated by \citet{BonaldiTrecs2019} at three integration times: 8, 100, and 1000 hours. All images are centred on the right ascension (RA) = $0^\circ$ and the declination (Dec) = $-30^\circ$. The full angular extent or the Field of View (FoV) of the map varies by frequency: $5.5^\circ$ at 560~MHz, $2.2^\circ$ at 1.4~GHz, and $0.33^\circ$ at 9.2~GHz. The corresponding pixel sizes ($\delta p$) are $0.60\arcsec$, $0.24\arcsec$, and $0.037\arcsec$, respectively, resulting in image dimensions of $32768 \times 32768$ pixels for all bands. A detailed description of the dataset and associated parameters is provided in \citet{SKAData2020}. For each map, the SKA SDC1 provides a true catalogue that lists the sources present in the full map. In addition, a training catalogue is provided that covers a small portion, 5\% of the full map, to test ML methods. Each source in the training catalogue is described by the following parameters that represent its position, morphology, and observational characteristics.
\begin{itemize}
    \item \texttt{ID}: Unique identifier for each source,
    \item \texttt{Right Ascension (RA) core, Declination (Dec) core}: Core position in degrees,
    \item \texttt{RA centroid, Dec centroid}: Centroid position in degrees,
    \item \texttt{x}, \texttt{y}: Coordinates of the centroid pixels (indexing starts from 0).
    \item \texttt{FLUX} (${f}$): Integrated flux density in Janskys (Jy),
    \item \texttt{Core frac} ($C_{\rm frac}$): Ratio of core flux to total flux (non-zero for AGNs only),
    \item \texttt{Bmaj} ($b_{\rm maj}$), \texttt{Bmin} ($b_{\rm min}$): Major and minor axes in arcseconds,
    \item \texttt{PA} ($\phi$): Position angle measured with respect to $b_{\rm maj}$ in degrees,
    \item \texttt{SIZE} ($C_{\rm profile}$): Source brightness profile class (1 = Largest Angular Size(LAS), 2 = Gaussian, 3 = Exponential),
    \item \texttt{CLASS} ($C_{\rm spectral}$): Source spectral class (1 = Steep-Spectrum AGN (SS-AGN), 2 = Flat-Spectrum AGN (FS-AGN), 3 = Star-Forming Galaxy (SFG)),
    \item \texttt{SELECTION}: Indicates whether the source is included in the simulated map due to noise level (1 = yes, 0 = no),
\end{itemize}

The SDC1 defines four main task categories, i.e., source detection, characterisation, spectral classification, and brightness profile classification.
\begin{enumerate}
    \item \textbf{Source Detection}: Identify and locate the RA and Dec centroids of radio sources. For AGNs, the core position should also be determined if possible.
    
    \item \textbf{Source Characterisation}: Estimate the following physical properties for each detected source:
    \begin{itemize}
        \item ${f}$
        \item ${b_{\rm maj}}$, ${b_{\rm min}}$
        \item $\phi$ with respect to ${b_{\rm maj}}$, measured clockwise from the west.
        \item ${C_{\rm frac}}$ (non-zero only for AGNs),
    \end{itemize}
    
    \item \textbf{Source Spectral Classification}: Classify each source into one of the following three categories on the basis of the spectral characteristics of the sources.
    \begin{itemize}
        \item SS AGN,
        \item FS AGN,
        \item SFG.
    \end{itemize}
    
    \item \textbf{Source Brightness Classification}: Classify the size profile of each source into one of the following three categories based on the source's brightness profile.
    \begin{itemize}
        \item LAS,
        \item Gaussian,
        \item Exponential.
    \end{itemize}
\end{enumerate}

\section{Catalogue Scoring Procedure} \label{Scoring_procedure}
In the SKA~SDC1 challenge, a true catalogue is provided, and a submitted (predicted) catalogue is generated using any statistical or supervised ML approach. These catalogues are then cross-matched to evaluate performance, following the procedures defined by \citet{Bonaldi_2020_10} and \citet{SKAData2020}. Due to extreme source crowding ($\sim 50$ sources arcmin$^{-2}$ at 560\, MHz), a custom two-stage cross-matching procedure was implemented.

\subsection{Binning of Catalogues}
To reduce computational complexity, both the True and Submitted catalogues were binned into 5 declination bins and 5 logarithmic flux density bins, resulting in 25 sub-catalogues. Cross-matching is performed within the corresponding bins only. The True catalogue bins are twice the size of the Submitted catalogue bins, allowing a half-bin overlap between adjacent True catalogue bins. This reduces the computational complexity by a factor of 12.5.

\subsection{Cross-Matching with Position, Flux, and Size}
The cross-match minimises a multi-dimensional distance $D$ between True and Submitted sources, which includes positional, flux, and size differences:
\begin{equation}
D = \sqrt{d_{pos}^2 + d_{flux}^2 + d_{size}^2}.
\end{equation}
Individual terms are defined as follows.
\begin{align}
d_{pos} &\propto \frac{\sqrt{(x-\bar{x}_{t})^2 + (y-\bar{y}_{t})^2}}{\bar{S}_{t}}, \\
d_{flux} &\propto \frac{|f-f_{t}|}{f_{t}}, \\
d_{size} &\propto \frac{|S-S_{t}|}{\bar{S}_{t}},
\end{align}
Here, the subscript $t$ denotes values from the true catalogue, $(x, y)$ are the pixel coordinates corresponding to RA and Dec, and $f$ is the integrated flux density. The average source size is defined as $S = (b_{\mathrm{maj}} + b_{\mathrm{min}})/2$. Each term is normalised such that a typical $1\sigma$ measurement error contributes $\sim 1$ to $D$, and only matches with $D < 5$ are accepted. The quantity $\bar{S}_t$ represents the largest source axis convolved with the synthesised beam, as observed by \citet{Cornu2024}. The cross-match is performed in two steps:
\begin{enumerate}
    \item A positional-only crossmatch is performed using the publicly available software package \citep{Riccio_2016}, retaining all candidate matches with $D_{pos}<3$.
    \item Among the candidate matches, the true source with minimum $D$ is selected as the match if $D<5$, resulting in one or no match per detection.
\end{enumerate}

\subsection{Null-Test Crossmatch}
To quantify chance matches due to source crowding, null-test catalogues are created by randomly repositioning the sources in the submitted catalogue while preserving their flux and size distributions. The same cross-match procedure is applied, and the number of chance matches $N_{\rm null}$ is subtracted when computing completeness and reliability.

\subsection{Completeness and Reliability}
Completeness $C$ and reliability $R$ are defined as
\begin{align}
C(\log f) &= \frac{N_{\mathrm{match}}(\log f) - N_{\mathrm{null}}(\log f)}{N_{\mathrm{true}}(\log f)}, \\
R(\log f_{\mathrm{appr}}) &= \frac{N_{\mathrm{match}}(\log f_{\mathrm{appr}}) - N_{\mathrm{null}}(\log f_{\mathrm{appr}})}{N_{\mathrm{det}}(\log f_{\mathrm{appr}})},
\end{align}
where $N_{\mathrm{match}}$ denotes the histogram of the sources matched between the true and submitted catalogues, $N_{\mathrm{true}}$ and $N_{\mathrm{det}}$ are the histograms of the true and submitted catalogues, respectively. Here, $f$ denotes the integrated flux density, while $f_{\mathrm{appr}}$ denotes the apparent integrated flux density.

\subsection{Overall Scoring}
The overall score accounts for $C$, $R$, and the accuracy of the estimated source property. For a Submitted catalogue at frequency $\nu$ and integration time $T$, the final score is
\begin{equation}
F_{\mathrm{sco}}(\nu) = \sum_{i=1}^{N_{\rm match}} w_i - \frac{N_{\rm false}\,(= N_{\rm det} - N_{\rm match})}{\rm FoV(\nu)},
\end{equation}
where $w_i$ is the weight per source, taking into account the precision of seven properties: position $(RA, Dec)$, $f$, $C_{\rm frac}$, $b_{\rm maj}$, $b_{\rm min}$, $\phi$, and $C_{\rm spectral}$.
For each property $j$, the weight per-source is
\begin{equation}
w_{i,j} = \frac{1}{7} \, \max\Big(1 - \frac{\epsilon_{i,j}}{\tau_j},0\Big),
\end{equation}
where $\epsilon_{i,j}$ is the error of the property $j$ for the source $i$, and $\tau_j$ is the threshold for that property as provided in \citet{Bonaldi_2019_SDC1_Scoring}. For the source spectral class (\(C_{\rm spectral}\)), $w_{i,j}=1/7$ if correctly classified, and 0 otherwise. The total weight per source is
\begin{equation}
w_i = \sum_{j=1}^7 w_{i,j},
\end{equation}
ensuring $0 \le w_i \le 1$.
The average source accuracy is then defined as
\begin{equation}
\bar{w} = \frac{1}{N_{\mathrm{match}}} \sum_{i=1}^{N_{\mathrm{match}}} w_i ,
\end{equation}
In this work, we focus on the first two categories: source detection and characterisation. Source detection yields the centre pixel coordinates $(X_c, Y_c)$, which are converted to global sky coordinates (RA, Dec), together with the width ($W$) and height ($H$) of the spatial extent surrounding the source and the associated confidence score ($C_{\mathrm{conf}}$). The characterisation model is used to estimate key physical parameters, namely $f$, $b_{\mathrm{maj}}$ and $b_{\mathrm{min}}$, and $\phi$. We do not consider the characterisation of $C_{\mathrm{frac}}$, $C_{\mathrm{spectral}}$, and $C_{\mathrm{profile}}$ in this study.

\section{Data Preprocessing} \label{Data_Preprocessing}
In this section, we describe the training data for the source detection and characterisation model. To explore our methods, we study the simulated map of 560 MHz, corresponding to 1000 hours of time integration. The training region is centered at RA $=-0.3362^{\circ}$ and Dec $=-29.6737^{\circ}$, covering approximately $0.68^{\circ}$ in RA and $0.53^{\circ}$ in Dec, corresponding to an image size of $4035 \times 3189$ pixels. The training catalogue contains 274,901 source entries. After filtering to include sources with \texttt{SELECTION} = 1 (i.e., sources injected into the simulated map due to noise level), approximately 190,000 sources are retained for training. The majority of sources are \(C_{\rm{spectral}}=3\) with \(C_{\rm{profile}}=2\), followed by \(C_{\rm{spectral}}=1\) with \(C_{\rm{profile}}=1\) and \(C_{\rm{spectral}}=2\) with \(C_{\rm{profile}}=2\) (see Table~\ref{tab:combined_tables}\(\ a\)).

The input images are provided in units of $\mathrm{Jy\,beam^{-1}}$ and are not corrected for the primary beam. Although a primary beam correction -- performed by multiplying each pixel by its corresponding beam gain value -- would convert the images into intrinsic surface brightness units (Jy), we intentionally avoided this correction. The noise in the simulated images is spatially uniform, whereas the primary beam response decreases radially outward from the centre. Applying a beam correction would amplify noise in the outer regions where the beam gain is low, degrading the signal-to-noise ratio (\(\rm SNR\)) and increasing the likelihood of false detections. This concern is particularly relevant for these simulations, which are based on a single pointing. Instead of applying the primary beam correction directly to the images, we incorporate its effect into the $f$ of each source in the training catalogue. Specifically, we interpolate the primary beam values corresponding to the central location of each source and multiply them by $f$ to obtain what we define as the \textit{apparent flux} (${f_{\rm appr}}$), following the approach described in \citet{Cornu2024}. In addition, we use ${f_{\rm appr}}$ as the primary selection criterion to determine which sources are included in the training process.

A major challenge in source detection is the wide dynamic range in spatial extent. Detecting both compact and extended sources with a single model is therefore non-trivial. Most sources have $b_{\mathrm{maj}}$ values in the range $0.1\arcsec \leq b_{\mathrm{maj}} < 2\arcsec$, while sources with $b_{\mathrm{maj}} \geq 2\arcsec$ are comparatively rare. As a result, the number of available training samples decreases rapidly with increasing $b_{\mathrm{maj}}$ (see Table~\ref{tab:BMAJ_sources_Anchors_Grid_size}).

To address this imbalance and improve detection performance across spatial scales, we develop two dedicated detection models: \texttt{Single-Scale Detection (SSD)} and \texttt{Multi-Scale Detection (MSD)}. In addition, we introduce a \texttt{Multi-Scale Characterisation (MSC)} model for estimating source physical parameters. The overall architectures of these models are shown in Figures~\ref{fig:model_architecture}, \ref{fig:model_architecture_M}, and \ref{fig:parameter_model}. The \texttt{SSD} model is optimised for detecting small sources defined in the range $0.1\arcsec < b_{\mathrm{maj}} < 6\arcsec$, while the \texttt{MSD} model targets medium and extended sources with $4\arcsec < b_{\mathrm{maj}} < 128\arcsec$. Following \citet{SKAData2020}, a source is classified as compact if $b_{\mathrm{maj}} < 3\,\delta p$ ($1.8\arcsec$ at 560~MHz), and extended otherwise. Using this criterion, $\sim95\%$ of the sources in the training set are compact (see Table~\ref{tab:combined_tables}\,(b)). Consequently, the \texttt{SSD} model is expected to detect a larger fraction of sources than the \texttt{MSD} model. The \texttt{MSC} model estimates the source parameters $(f_{\mathrm{appr}}, b_{\mathrm{maj}}, b_{\mathrm{min}}, \sin\phi, \cos\phi)$ using the bounding boxes of the sources over the full range of size $0.1\arcsec < b_{\mathrm{maj}} < 128\arcsec$. A bounding box (bbox) spatially encloses a source and is defined by its centre coordinates together with its width and height. This framework comprises \texttt{SSD} for compact-source detection, \texttt{MSD} for medium- and large-source detection, and \texttt{MSC} for source characterisation. Together, these components enable automated detection and physical characterisation of radio sources over a wide range of source sizes. Further details of the model architectures are provided in Section~\ref{detectors}.
\begin{table}
\centering
\caption{Distribution of training sources by \(C_{\rm profile}\), \(C_{\rm spectral}\), compactness, and extended.}
\label{tab:combined_tables}
\textbf{(a) $\mathbf{C_{\mathrm{profile}}}$ vs $\mathbf{C_{\mathrm{spectral}}}$} \\[0.5pt]
\begin{tabular}{c|cccc}
\hline 
& $\mathbf{C_{\mathrm{spectral}}=1}$ & $\mathbf{C_{\mathrm{spectral}}=2}$ & $\mathbf{C_{\mathrm{spectral}}=3}$ & \textbf{Total} \\
\hline
$\mathbf{C_{\mathrm{profile}}=1}$          & 6127 & --    & --      & 6127 \\
$\mathbf{C_{\mathrm{profile}}=2}$     & 380  & 3728  & 179946  & 184054 \\
$\mathbf{C_{\mathrm{profile}}=3}$  & --   & --    & 372     & 372 \\
\textbf{Total} & 6507 & 3728 & 180318 & 190553 \\
\hline
\end{tabular}

\vspace{0.5em}

\textbf{(b) $\mathbf{C_{\mathrm{spectral}}}$ vs $\mathbf{Compact}$ ($\mathbf{b_{\mathrm{maj}} \leq 1.8\arcsec}$) and $\mathbf{Extended}$ ($\mathbf{b_{\mathrm{maj}} \geq 1.8\arcsec}$) sources} \\[1pt]

\begin{tabular}{c|ccc}
\hline
& \textbf{Compact} & \textbf{Extended} & \textbf{Total} \\
\hline
$\mathbf{C_{\mathrm{profile}}=1}$ & 340     & 6167   & 6507 \\
$\mathbf{C_{\mathrm{profile}}=2}$ & 3718    & 10     & 3728 \\
$\mathbf{C_{\mathrm{profile}}=3}$ & 176783  & 3535   & 180318 \\
\textbf{Total} & 180841 & 9712 & 190553 \\
\hline
\end{tabular}
\end{table}

\subsection{{Ground-truth bounding box generation}}
 \label{Bounding_Box}
 Here, we describe the procedure used to generate gt-bboxes for training. First, the sky coordinates (RA and Dec centroids) are converted to pixel coordinates \((X_c, Y_c)\) using the World Coordinate System (WCS) transformation \citep{astropy_2013}. The width \(W\) and height \(H\) of the gt-bboxes are then derived from the source parameters \(b_{\mathrm{maj}}\), \(b_{\mathrm{min}}\), \(\phi\), \(\delta p\), \(C_{\mathrm{profile}}\), and \(C_{\mathrm{spectral}}\). The full width at half maximum (FWHM) along the major and minor axes is denoted by \(w_1\) and \(w_2\), respectively, and is estimated from these physical properties described below:
 
 For \(C_{\mathrm{spectral}} = 1\), sources modelled with either \(C_{\mathrm{profile}} = 1\) or \(2\) adopt \(w_1 = b_{\mathrm{maj}}/2\) and \(w_2 = b_{\mathrm{min}}/2\). All sources with \(C_{\mathrm{spectral}} = 2\) are modelled with a circular profile (\(C_{\mathrm{profile}} = 2\)), for which \(w_1 = w_2 = b_{\mathrm{maj}} = b_{\mathrm{min}}\). For \(C_{\mathrm{spectral}} = 3\), most compact sources are modelled with \(C_{\mathrm{profile}} = 2\), using \(w_1 = b_{\mathrm{maj}}\) and \(w_2 = b_{\mathrm{min}}\), while a small fraction (0.2\%) is modelled with \(C_{\mathrm{profile}} = 3\) (see Table~\ref{tab:combined_tables}), for which \(w_1 = \sqrt{2}\,b_{\mathrm{maj}}\) and \(w_2 = \sqrt{2}\,b_{\mathrm{min}}\). For extended sources with \(C_{\mathrm{spectral}} = 3\), those modelled with \(C_{\mathrm{profile}} = 2\) use \(w_1 = b_{\mathrm{maj}}/\sqrt{2}\) and \(w_2 = b_{\mathrm{min}}/\sqrt{2}\), while those with \(C_{\mathrm{profile}} = 3\) use \(w_1 = b_{\mathrm{maj}}\) and \(w_2 = b_{\mathrm{min}}\). To ensure well-defined gt-bboxes for unresolved or nearly point-like sources, a regularisation term \((2\,\delta p)^2\) is added, ensuring a non-zero gt-bbox size
below the resolution limit. The effective FWHM values are then defined as
\[
b_1 = \sqrt{w_1^2 + (2\,\delta p)^2}, \qquad
b_2 = \sqrt{w_2^2 + (2\,\delta p)^2}.
\]
Using \(\phi\), measured with respect to \(b_{\rm{maj}}\), the
rotated projections of the source ellipse with respect to \(b_{\rm{maj}}\), \(b_{\rm{min}}\) are computed. The projection due to \(b_{\rm{min}}\) is
obtained by applying an additional phase shift of \(\pi/2\):
\[
x_1 = \frac{b_1}{\delta p}\cos\phi, \qquad
x_2 = \frac{b_2}{\delta p}\cos\!\left(\phi + \frac{\pi}{2}\right),
\]
\[
y_1 = \frac{b_1}{\delta p}\sin\phi, \qquad
y_2 = \frac{b_2}{\delta p}\sin\!\left(\phi + \frac{\pi}{2}\right).
\]
The final width \(W\) and height \(H\) of the gt-bbox are then given by
\[
W = 2.5\,\sqrt{x_1^2 + x_2^2}, \qquad
H = 2.5\,\sqrt{y_1^2 + y_2^2},
\]
To conservatively encompass the full source brightness, we apply a scaling factor of 2.5, which corresponds approximately to the number of pixels across the synthesised beam. The resulting gt-bbox for each source in the \(512 \times 512\) pixel image is defined as
\begin{equation}
\mathrm{bbox} = (X_c, Y_c, W, H),
\label{eq:BBox}
\end{equation}
with \(X_c\) and \(Y_c\) denoting the source centre coordinates in the image plane.

\subsection{Detection Data Preprocessing}
\begin{figure}
    \centering
    \includegraphics[width=.4\textwidth]{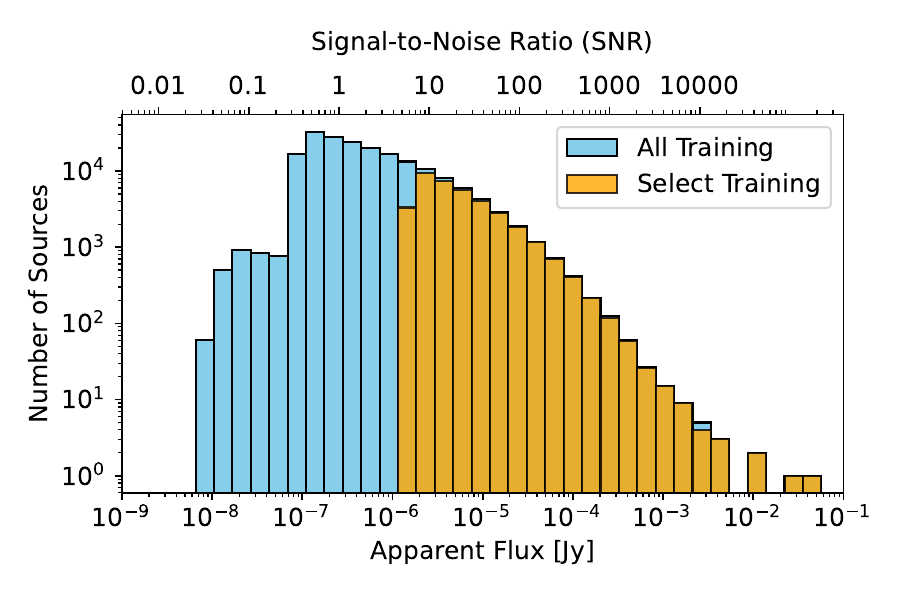}
    \caption{Log--log histograms of the apparent integrated flux density \(f_{\mathrm{appr}}\) for the full training catalogue and the selected training sources. The upper axis indicates the corresponding SNR for \(\sigma = 0.255 \times 10^{-6}\,\mathrm{Jy}\). All selected sources lie above \(8\sigma\). Source were selected based on the equation \eqref{eq:flux_threshold}}.
    \label{fig:Flux_Histogram_Log_SNR}
\end{figure}

To prepare the training data for the detection models, we first discarded faint and highly unresolved sources with \(f_{\mathrm{appr}} < 1.6 \times 10^{-6}\,\mathrm{Jy}\), \(b_{\mathrm{maj}} < 0.1\arcsec\), and \(b_{\mathrm{min}} \leq 0.01\arcsec\). We further retained only sources with sufficient surface brightness relative to their spatial extent by applying the following criterion:
\begin{equation}
\frac{f_{\mathrm{appr}}}{(b_{\mathrm{maj}}\, b_{\mathrm{min}})/(\delta p)^2}
\geq 5 \times 10^{-8}\,\mathrm{Jy\,pixel^{-2}} .
\label{eq:flux_threshold}
\end{equation}
The thermal noise of the map is \(\sigma = 0.255 \times 10^{-6}\,\mathrm{Jy\,beam^{-1}}\), as reported by \citet{Bonaldi_2020_10}. The \(\rm SNR\) of the selected training sources was calculated using the full training catalogue (see Fig.~\ref{fig:Flux_Histogram_Log_SNR}). All retained sources have \(\mathrm{SNR} > 8\sigma\).

To generate training samples for both detection models, \texttt{SSD} and \texttt{MSD}, we uniformly sampled grid points in RA and Dec within the training region. For \texttt{SSD}, a \(12 \times 10\) grid was used, as illustrated in the right panel of Fig.~\ref{fig:full_and_zoom}, which yielded 120 training samples. For \texttt{MSD}, a denser grid \(16 \times 14\) was adopted to increase the number of training images, as medium and large sources are comparatively rare, resulting in 224 training samples. Each point in the grid defines the centre of a \(309.5\arcsec \times 309.5\arcsec\) FITS cutout \citep{FITS}, corresponding to a \(512 \times 512\) pixel image. To increase data diversity, each training image was augmented by rotations of \(90^\circ\), \(180^\circ\), and \(270^\circ\), as well as horizontal flipping. This expanded the datasets from 120 to 960 samples for the \texttt{SSD} and from 224 to 1792 samples for the \texttt{MSD}, improving model generalisability. In addition, 32 noise-only patches were extracted from the noise regions of the full map, indicated by the white boxes on the full map on the left panel of Fig.~\ref{fig:full_and_zoom}. These were included to improve robustness against background noise, increasing the final training sets to 992 and 1824 samples for the \texttt{SSD} and \texttt{MSD}, respectively. Before being passed to the detection models, all images were normalised to the range \([0,1]\) using
\begin{equation}
I_d = \max\!\left(0,\,
\tanh\!\left(
\alpha \,
\frac{I_0 - I_{\min}}{I_{\max} - I_{\min}}
\right)
\right),
\label{eq:Det_Normalization}
\end{equation}
where \(I_0\) denotes the original pixel brightness in \(\mathrm{Jy\,beam^{-1}}\) and \(I_d\) the normalised pixel values used by the detectors. The clipping limits are set to \(I_{\min} = 4 \times 10^{-7}\,\mathrm{Jy\,beam^{-1}}\) and \(I_{\max} = 4 \times 10^{-5}\,\mathrm{Jy\,beam^{-1}}\), while \(\alpha = 3\) controls the slope of the transformation. This normalisation enhances faint emission relative to the background while compressing bright pixels associated with extended sources \citep{Cornu2024}. As the detector is optimised for source localisation rather than precise parameter estimation, this normalisation is appropriate only for detection tasks. An example of a normalised training image with source gt-bboxes is shown in Fig.~\ref{fig:A_training_example}. Highly blended training sources superposed on bright emission were also discarded using non-maximum suppression (NMS) \citep{nms2017} with an intersection-over-union (IoU) threshold of 0.2 based on  \(f_{\mathrm{appr}}\). The NMS procedure relies on the IoU metric,
\begin{equation}
\mathrm{IoU}(\mathbf{a}, \mathbf{b}) =
\frac{|\mathbf{a} \cap \mathbf{b}|}{|\mathbf{a} \cup \mathbf{b}|}.
\label{eq:IOU_def}
\end{equation}

\subsection{Detector input and output encoding} \label{Data_encoding}
Both detection models adopt a YOLOv3-style formulation. The input is a \(512 \times 512\times1\) image, normalised using Eq.~\ref{eq:Det_Normalization}, and the corresponding output label is a tensor of shape \(G \times G \times A \times 5\). Here, \(G\) denotes the number of grid cells along each image axis, and \(A\) is the number of anchors per grid cell. The output tensor is initialised to zero. The image is partitioned into a uniform \(G\times G\) grid, where each grid cell spans
\begin{equation}
\Delta = \frac{512}{G}
\end{equation}
pixels along both axes. A gt-bbox is assigned to a grid cell indexed by the integer pair \((g_x, g_y)\), obtained from the coordinates of the central pixel \((X_c, Y_c)\) of the source as
\begin{equation}
x_c = \frac{X_c}{\Delta}, \quad
y_c = \frac{Y_c}{\Delta},
\end{equation}
\begin{equation}
g_x = \mathrm{int}(x_c), \quad
g_y = \mathrm{int}(y_c),
\end{equation}
where \(\mathrm{int}(\cdot)\) denotes conversion to an integer grid index. The quantities \(x_c\) and \(y_c\) represent the centre of the gt-bbox expressed in units of the grid and continuously vary in the range \([0, G)\). The \(W\) and \(H\) of the gt-bbox, expressed in pixel units, are encoded linearly as
\begin{equation}
w = \frac{W}{\Delta}, \quad
h = \frac{H}{\Delta}.
\end{equation}

Within each cell of the grid, all \(A\) anchors are initially set to zero. For a given gt-bbox, the IoU is computed with all anchors associated with the corresponding grid cell. The anchor that yields the maximum IoU is selected as the positive anchor, for which \(C_{\mathrm{conf}}\) is set to 1 and the parameters \((x_c, y_c, w, h)\) are assigned. All remaining anchors retain zero values and are treated as negative anchors (i.e., the background). The loss is calculated between the encoded ground-truth parameters \((C_{\mathrm{conf}}, x_c, y_c, w, h)\) and the corresponding network predictions \((C'_{\mathrm{conf}}, x'_c, y'_c, w', h')\).

During inference, the predicted bounding-box parameters are converted back to pixel units by multiplying by the grid-cell size \(\Delta\):
\begin{align}
X'_{c} &= x'_{c}\times\Delta, &
Y'_{c} &= y'_{c}\times\Delta, \\
W' &= w'\times\Delta, &
H' &= h'\times\Delta.
\end{align}

\begin{figure*}
    \centering
    \includegraphics[width=\textwidth]{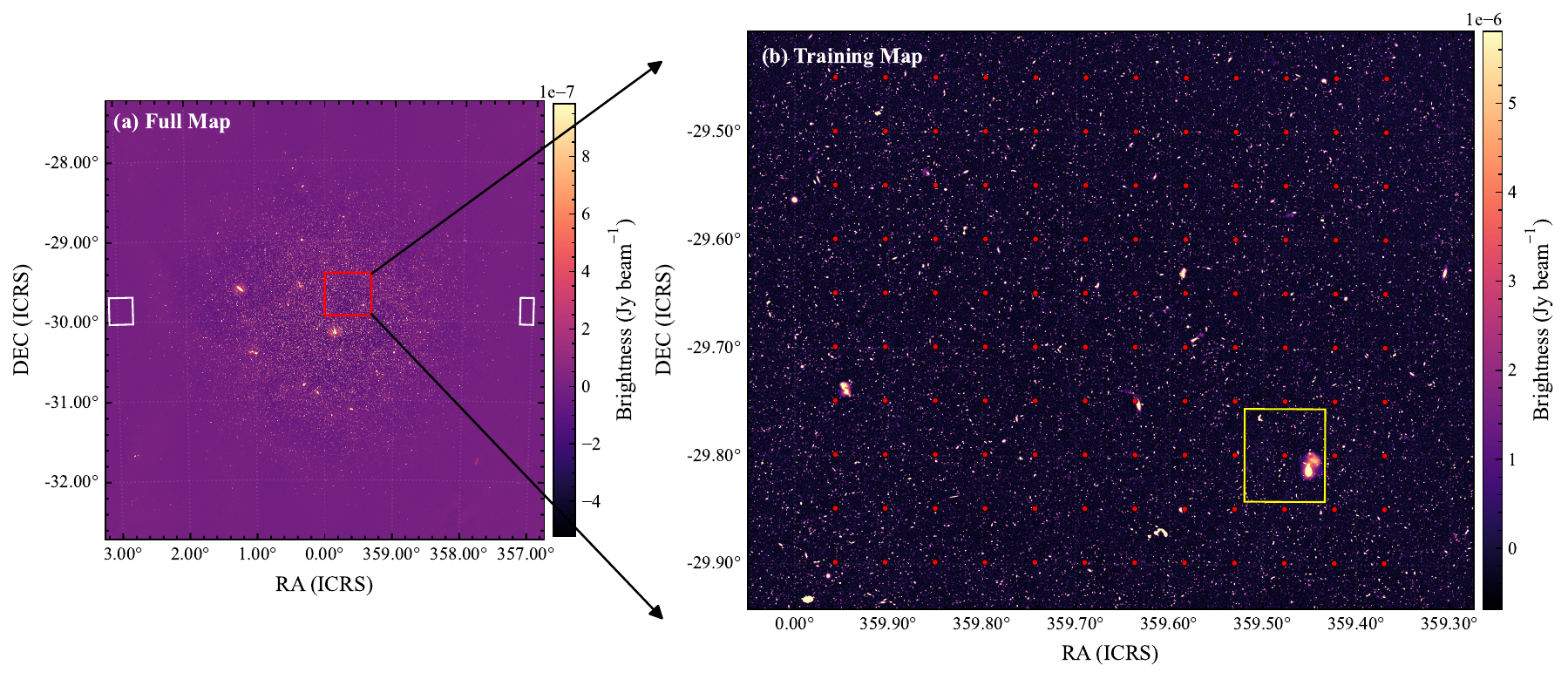}
    \caption{Visualisation of the full map (left) and the corresponding training region (right). The red box in the left panel marks the training region, with arrows indicating the correspondence to the zoomed-in view. White boxes denote background or noise regions. Both images are shown in pixel units with brightness in Jy\,beam$^{-1}$ in the ICRS frame. The yellow box in the right panel highlights a \(309.5\arcsec \times 309.5\arcsec\) (\(512 \times 512\) pixel) training cutout as shown in the Fig.~\ref{fig:A_training_example}.}
    \label{fig:full_and_zoom}
\end{figure*}

\begin{figure}
\centering  \includegraphics[width=\columnwidth]{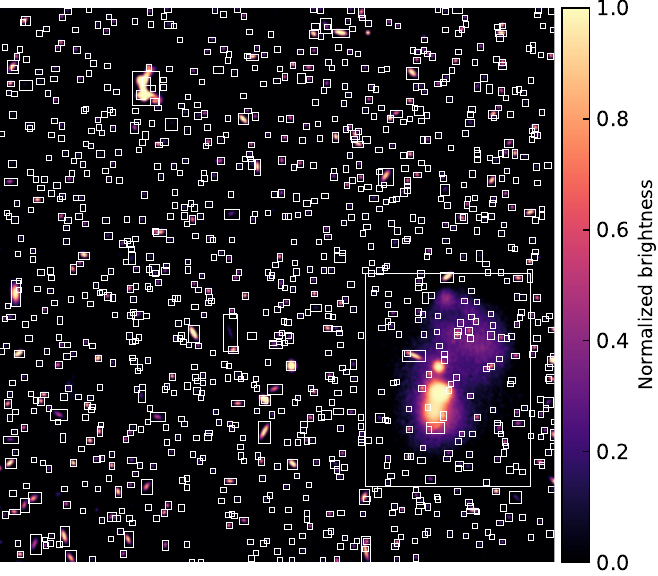}
\caption{Example cutout from the training region, centred at RA $= -0.5195^\circ$ and Dec $= -29.7991^\circ$ (highlighted in yellow in the right panel of Fig.~\ref{fig:full_and_zoom}), normalized for the detection task using equation~\ref{eq:Det_Normalization}. gt-bboxes are shown for compact, medium, and extended sources spanning $0.1\arcsec \leq b_{\rm maj} < 200\arcsec$.}
\label{fig:A_training_example}
\end{figure}

This encoding–decoding scheme preserves a direct correspondence between the detector output and the physical pixel scale. For the \texttt{SSD} model, each \(512 \times 512\times1\) image is divided into a \(64 \times 64\) grid (\(\Delta = 8\) pixels), with \(A=9\) anchors per cell, producing an output tensor of shape \((64, 64, 9, 5)\). The fine grid resolution makes the \texttt{SSD} model well suited for detecting compact sources. Figure~\ref{fig:full_and_zoom} shows the full map and training region, while Fig.~\ref{fig:A_training_example} presents an example training cutout with gt-bbox for each source spanning \(0.1\arcsec \leq b_{\mathrm{maj}} < 128\arcsec\). In contrast, the \texttt{MSD} model operates on two coarser grid scales to detect medium and large sources. Medium-sized sources (\(b_{\mathrm{maj}} = 4\arcsec\)–\(40\arcsec\)) are detected using a \(32 \times 32\) grid (\(\Delta = 16\) pixels), producing a \((32, 32, 9, 5)\) tensor. Large sources (\(b_{\mathrm{maj}} = 30\arcsec\)–\(128\arcsec\)) are detected using a \(16 \times 16\) grid (\(\Delta = 32\) pixels), yielding a \((16, 16, 9, 5)\) tensor. The use of multiple grid scales enables effective detection across a broad range of spatial extents.

The \texttt{SSD} model produces a total of 36864 predictions, known as S-boxes, at the output layer \(O\_{11}\) for a \(512 \times 512\) pixel image (Fig.~\ref{fig:model_architecture}). Similarly, the MSD model produces 9216 M-boxes and 2304 L-boxes, as shown in Fig.~\ref{fig:model_architecture_M}.

Anchor boxes serve as initial shape priors for detection, with the network predicting offsets to refine $w'$ and $h'$. To derive suitable anchor shapes, we apply a \(k\)-means clustering algorithm \citep{Jin2010} to the width and height of the gt-bboxes  \((W, H)\), using IoU as distance metric. Clustering is performed separately for different \(b_{\mathrm{maj}}\) ranges. The resulting anchor configurations, together with the corresponding grid resolutions and detection models, are summarised in Table~\ref{tab:BMAJ_sources_Anchors_Grid_size}.

\begin{table}
\centering
\caption{Distribution of training sources binned by \(b_{\rm maj}\), with 3 anchors (A) per bin, and grid resolutions (G) used for training in the respective detection models. $N$ is the number of training sources in the corresponding bin. All sources have \(f_{\rm appr} \geq 1.6\times 10^{-6} \, \text{Jy}\).}
\scriptsize
\setlength{\tabcolsep}{14pt}
\begin{tabular}{ccccc}
\toprule
\textbf{\(b_{\rm maj}\) (\(\arcsec\))} & \textbf{N} & \textbf{A} & \textbf{G} & \textbf{Model} \\
\midrule
$0.1 < b_{\rm maj} \leq 1$   & 30641 & A1--A3 & 64 & SSD \\
$1 < b_{\rm maj} \leq 3$      & 7624  & A4--A6 & 64 & SSD \\
$3 < b_{\rm maj} \leq 6$      & 1162  & A7--A9 & 64 & SSD \\
$3 < b_{\rm maj} \leq 10$     & 1413  & A10--A12 & 32 & MSD \\
$10 < b_{\rm maj} \leq 20$    & 173   & A13--A15 & 32 & MSD \\
$20 < b_{\rm maj} \leq 40$    & 98    & A16--A18 & 32 & MSD \\
$30 < b_{\rm maj} \leq 60$    & 73    & A19--A21 & 16 & MSD \\
$50 < b_{\rm maj} \leq 100$   & 68    & A22--A24 & 16 & MSD \\
$90 < b_{\rm maj} \leq 130$   & 18    & A25--A27 & 16 & MSD \\
\bottomrule
\end{tabular}
\label{tab:BMAJ_sources_Anchors_Grid_size}
\end{table}
\begin{figure}
    \centering
    \includegraphics[width=\columnwidth]{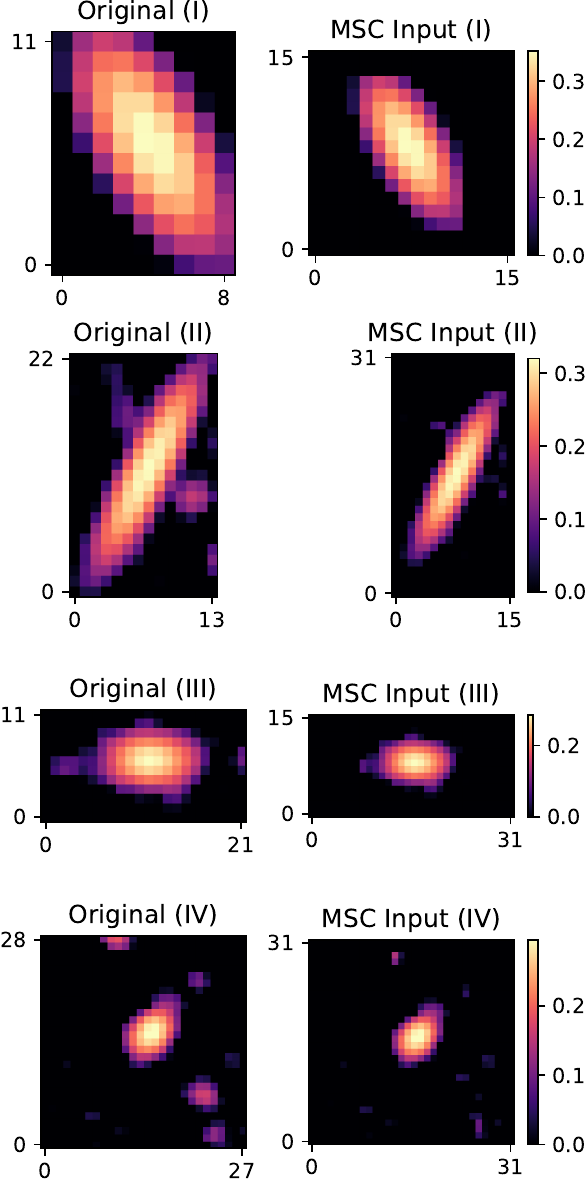}
    \caption{Some examples of gt-bboxes with varying sizes (left) and their padded versions used as \texttt{MSC} inputs (right): (I) $16\times16$, (II) $16\times32$, (III) $32\times16$, and (IV) $32\times32$, following normalisation using equation~\ref{eq:ch_log_norm_unit}.}
    \label{fig:dataset_examples}
\end{figure}

\subsection{Data Pre-processing For MSC}  \label{Char_Data_Preprocessing}
To prepare the inputs for the \texttt{MSC} model, the original brightness values (\(\mathrm{I_o}\)) are normalised using a clipped dynamic range defined by \(\mathrm{I_{min}} = 4 \times 10^{-7}\ \mathrm{Jy\ beam^{-1}}\) and \(\mathrm{I_{max}} = 0.5\ \mathrm{Jy\ beam^{-1}}\), where \(\mathrm{I_{max}}\) corresponds to the maximum pixel value in the image. Notably, this upper limit is four orders of magnitude higher than that used for detector normalisation (Eq.~\ref{eq:Det_Normalization}), with base-10 logarithmic normalisation to preserve relative brightness information as:
\begin{equation}
\mathrm{I_c}=\frac{\max\!\left(0,\,\log_{10}\!\left(\frac{\mathrm{I_o}}{\mathrm{I_{min}}}\right)\right)}{\log_{10}\!\left(\frac{\mathrm{I_{max}}}{\mathrm{I_{min}}}\right)}
\label{eq:ch_log_norm_unit}
\end{equation}
This normalisation constrains the values of \(\mathrm{I_c}\) to the range \([0,1]\), compresses the dynamic range of pixel brightness, and preserves structural and morphological information, thus improving the accuracy of parameter estimation. To further refine the training set, we apply a surface-brightness selection criterion and retain only sources satisfying
\begin{equation}
\frac{f_{\mathrm{appr}}}{W \times H}
>
5 \times 10^{-8}\ \mathrm{Jy\ pixel^{-2}},
\label{eq:flux_threshold_Char}
\end{equation}
which removes very low surface-brightness sources that are dominated by noise and contribute little to the training process.

Training samples for the \texttt{MSC} model are generated using the same images as used for the \texttt{SSD} model. The corresponding target labels for each input are the source parameters \(f_{\mathrm{appr}},\ b_{\mathrm{maj}},\ b_{\mathrm{min}},\ \sin(\phi)\) and \(\cos(\phi)\), each normalised to the range \([0,1]\). For sources below the resolution limit, \(\phi\) estimates become unreliable; therefore, for sources with \(b_{\mathrm{maj}} \leq 1\arcsec\), \(\phi\) is set to zero for training data. A key challenge arises in dense or crowded regions, where brightness from neighbouring sources can overlap with that of the target source and degrade model performance. To mitigate this effect, nearby sources within the target source gt-bbox are identified based on their Euclidean separation and masked using their respective gt-bboxes. To enable efficient batch processing, all target source regions extracted from the gt-bboxes are then reshaped to standardised dimensions by applying zero padding to the surrounding background, without altering the intrinsic source brightness.
The \texttt{MSC} model therefore operates on a discrete set of standardised input sizes \((W_{\mathrm{in}} \times H_{\mathrm{in}})\), namely \(16\times16\), \(16\times32\), \(32\times16\), \(32\times32\), \(32\times64\), \(64\times32\), \(64\times64\), \(64\times128\), \(128\times64\), and \(128\times128\), with the mapping from the gt-bbox dimensions \((W,H)\) to each input size given in Table~\ref{tab:uniform_w_h}.

\begin{table}
\centering
\caption{
Final adjusted bbox input width (\(W_{\rm in}\)) and height (\(H_{\rm in}\)) for the MSC model, together with the corresponding gt-bbox extent (\(W\), \(H\)). \(N\) denotes the number of sources within each spatial range, with the associated \(b_{\rm maj}\) interval.
}
\label{tab:uniform_w_h}
\setlength{\tabcolsep}{3pt}
\begin{tabular}{ccccc}
\toprule
\(W\) (pix) & \(H\) (pix) & \(W_{\rm in} \times H_{\rm in}\) & \(b_{1}\arcsec - b_{2}\arcsec\) & \(N\) \\
\midrule
$0$--$16$  & $0$--$16$   & $16\times16$   & \(0.01\arcsec - 8.90\arcsec\)     & 37612 \\
$0$--$16$  & $16$--$32$  & $16\times32$   & \(4.06\arcsec - 13.48\arcsec\) & 94 \\
$16$--$32$ & $0$--$16$   & $32\times16$   & \(3.51\arcsec - 13.67\arcsec\)  & 154 \\
$16$--$32$ & $16$--$32$  & $32\times32$   & \(3.77\arcsec - 17.70\arcsec\) & 74 \\
$0$--$32$  & $32$--$64$  & $32\times64$   & \(7.55\arcsec - 19.58\arcsec\)  & 30 \\
$32$--$64$ & $0$--$32$   & $64\times32$   & \(7.73\arcsec - 19.79\arcsec\) & 21 \\
$32$--$64$ & $32$--$64$  & $64\times64$   & \(11.27\arcsec - 30.55\arcsec\) & 12 \\
$0$--$64$  & $64$--$250$ & $64\times128$  & \(20.82\arcsec - 30.15\arcsec\) & 4 \\
$64$--$250$& $0$--$64$ & $128\times64$  & \(31.50\arcsec - 52.05\arcsec\) & 3 \\
$64$--$250$& $64$--$250$ & $128\times128$ & \(37.93\arcsec - 47.50\arcsec\) & 6 \\
\bottomrule
\end{tabular}
\end{table}

\section{Detectors Architecture and Methodology} \label{detectors}
\begin{figure}
    \centering
    \includegraphics[width=\linewidth]{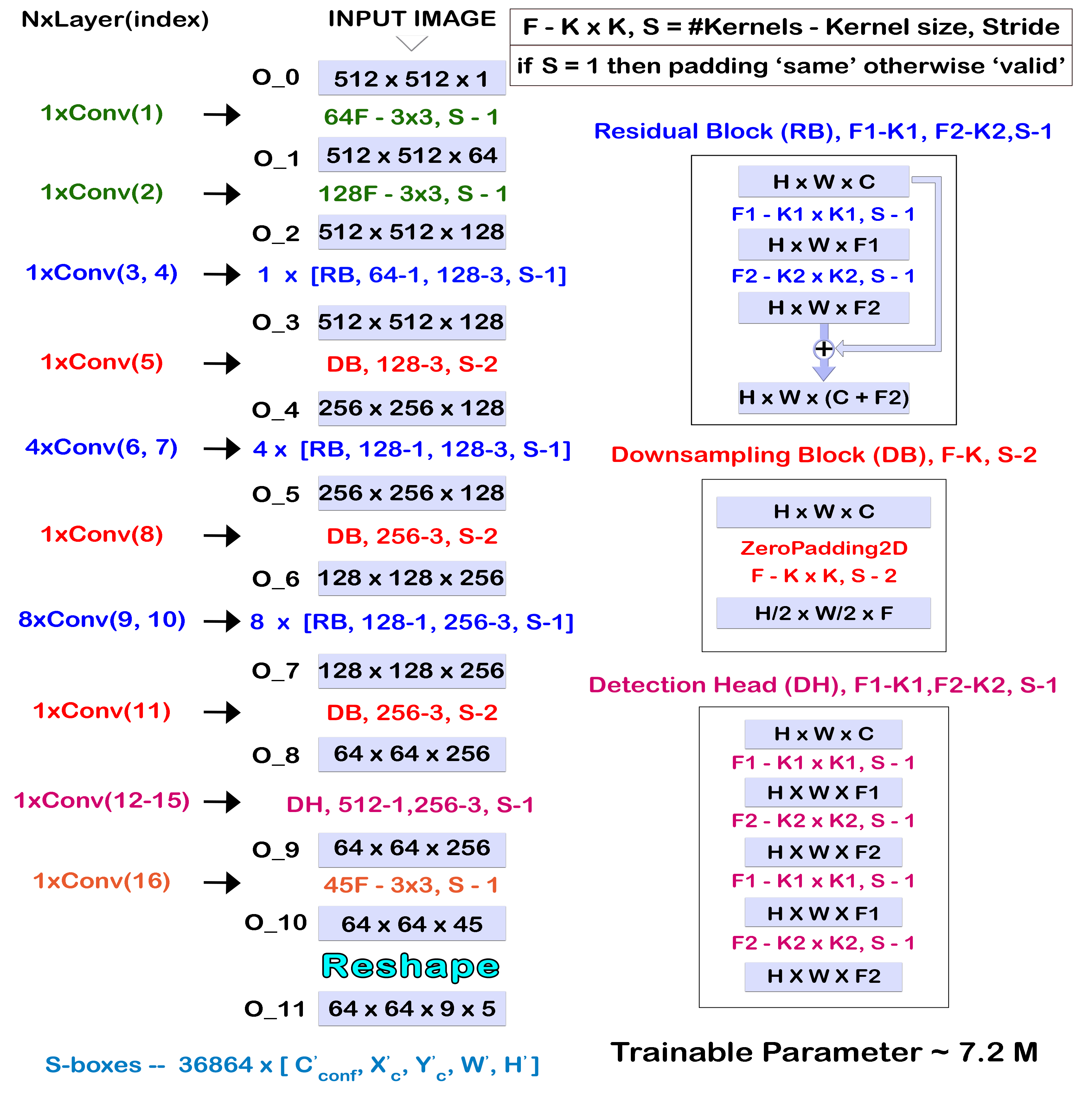}
    \caption{Architecture of the proposed \texttt{SSD} model. The model contains approximately 7.2 million trainable parameters. The network consists of residual blocks (blue), downsampling blocks (red), and a detection head (magenta). The number of filters (\(F\)), kernel sizes (\(K\)), and strides (\(S\)) are indicated.}
    \label{fig:model_architecture}
\end{figure}

\begin{figure}
    \centering
\includegraphics[width=1.\linewidth]{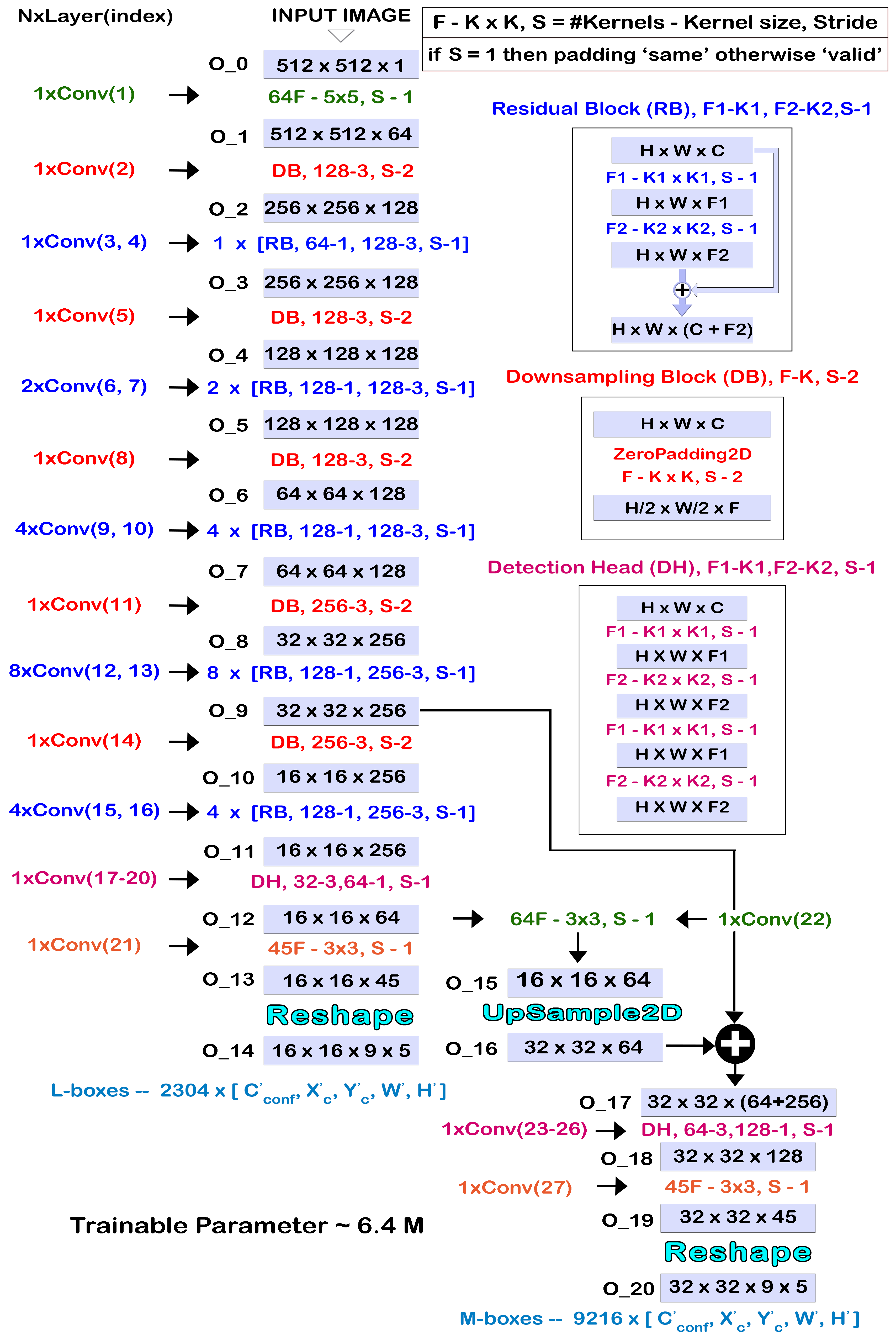}
\caption{Architecture of the proposed MSD model. The network includes residual blocks (blue), downsampling blocks (red), and a detection head (magenta), with the number of filters, kernel sizes, and stride values indicated, and contains approximately 6.4 million trainable parameters. The L-boxes ($O_{14}$) generate 2,304 predictions, while the M-boxes ($O_{20}$) produce 9,216 predictions, corresponding to large and medium source detections, respectively.}
\label{fig:model_architecture_M}
\end{figure}

This section describes the architectures of the \texttt{SSD} and \texttt{MSD} detection models. The \texttt{SSD} and \texttt{MSD} models contain approximately 7.2 million and 6.4 million trainable parameters, respectively. Their architectures are shown in Figures~\ref{fig:model_architecture} and~\ref{fig:model_architecture_M}. For an input brightness map \(I{(j,k)}\) and a kernel with \(\Theta{(l,m)}\) pixel values, the resulting output feature map $\tilde{I}$ with a pixel brightness value at $j, k$ is given by
\begin{equation}
\tilde{I}(j, k) = \tilde{A}\!\left(\sum_{l}\sum_{m} I{(j+l,\,k+m)}\,\Theta{(l,m)}\right),
\end{equation}
More details can be found in \citet{Bishop2024CNN}. 

In the \texttt{SSD} model, the first two convolutional layers (Conv(1) and Conv(2)) employ 64 and 128 kernels, respectively, with a kernel size of \(3 \times 3\) and a stride of 1. These layers extract low-level features while preserving the image's spatial resolution of \(512 \times 512\) pixels. In contrast, the \texttt{MSD} model uses a single initial convolutional layer (Conv(1)) with 64 kernels, a larger kernel size of \(5 \times 5\), and a stride of 1. This design reflects the fact that the \texttt{MSD} model primarily targets medium- and large-scale sources, for which fine-scale feature extraction is less critical. Following these initial layers, both networks apply a sequence of residual blocks and downsampling blocks to progressively extract higher-level features.

\subsection{Residual block}
Each residual block (RB) consists of two convolutional layers: a \(1 \times 1\) convolution followed by a \(3 \times 3\) convolution. The \(1 \times 1\) layer performs an efficient channel-wise feature transformation, enabling the network to capture fine-scale source details while preserving spatial resolution. 

Residual blocks learn a residual function rather than a direct mapping, which improves gradient propagation and mitigates the vanishing-gradient problem \citep{he_2015_ResNet}. The residual mapping can be written as
\begin{equation}
I_{\mathrm{out}} = \mathcal{R}(I_{\mathrm{in}}; \mathbf{\Theta}) + I_{\mathrm{in}},
\end{equation}
where \(I_{\mathrm{in}}\) and \(I_{\mathrm{out}}\) denote the input and output feature maps, respectively, and \(\mathcal{R}(I_{\mathrm{in}}; \mathbf{\Theta})\) represents the residual function learnt by the convolutional layers with trainable parameters \(\mathbf{\Theta}\).

\subsection{Downsampling Block}
A downsampling block (DB) first applies zero-padding to maintain alignment, followed by a $3 \times 3$ convolution with stride 2 and \textit{valid} padding. This reduces the spatial dimensions of the output half of the input dimension.

In general, the \texttt{SSD} uses 3 RBs and 3 DBs, whereas the \texttt{MSD} uses 5 RBs and 5 DBs (Figs.~\ref{fig:model_architecture} \& \ref{fig:model_architecture_M}).

\subsection{Detection Head}
While RBs and DBs extract hierarchical features, the Detection Head (DH) refines these features for the final detection task. DH forms the final stage of the network defined by Conv(12–15) in the \texttt{SSD} model and in \texttt{MSD}, Conv(23–26) for medium sources and Conv(17–20) for large sources. In the DH block, we employed four CNN layers, alternating between $1 \times 1$ and $3 \times 3$ kernel sizes while maintaining the same spatial dimensions, progressively refining the extracted features and mapping the spatial extent for source detection.

In the \texttt{MSD} model, to enhance medium-source detection, a Conv(22) is applied to $O_{12}$. The output $O_{15}$ is upsampled by linear interpolation and concatenated with $O_{9}$ along the channel dimension, producing $O_{17}$ of size $32 \times 32 \times 320$. Being a multi-scale network, information from different scales is combined to improve detection performance. Feature maps $O_{9}$ and $O_{11}$ detect medium and large sources, respectively. For large sources, the DH block (Conv(17–20)) is applied to $O_{11}$, followed by Conv(21) to generate predictions reshaped into a $16 \times 16 \times 9 \times 5$ tensor.  

All convolutional layers use Leaky ReLU \citep{xu2015LeakyRelu} with $\alpha = 0.1$, except the final layers (Conv(16) in \texttt{SSD} and Conv(21), Conv(27) in \texttt{MSD}). The activation is
\begin{equation}
\tilde{A}(x) =
\begin{cases}
x, & \text{if } x \geq 0, \\
\alpha x, & \text{if } x < 0,
\end{cases}
\end{equation}
where $\alpha$ is 0.1. Leaky ReLU prevents the ``dying ReLU'' problem by allowing a small gradient for negative inputs. Batch normalisation is not used, and all weights are initialised with a Glorot uniform initialisation \citep{glorot10a}.

The predictions which are reshaped to $64 \times 64 \times 9 \times 5$ for \texttt{SSD} and to $32 \times 32 \times 9 \times 5$ and $16 \times 16 \times 9 \times 5$ for \texttt{MSD}, where each tensor contains:
\begin{itemize}
    \item 9 anchors per grid cell,
    \item 5 values per anchor: raw outputs $({\hat{c}_{conf}}, \hat{x}_c, \hat{y}_c, \hat{w}, \hat{h})$, remapped to $(C'_{\rm conf}, x'_c, y'_c, w', h')$ for loss computation and final prediction format $(C'_{\rm conf}, X'_c, Y'_c, W', H')$ for inference as described in Section~\ref{Data_encoding}.
\end{itemize}

\section{Multi-Scale Characterisation (MSC) Model} \label{Multi_Scale_Characterization}
\begin{figure}
    \centering
    \includegraphics[width=0.9\linewidth]{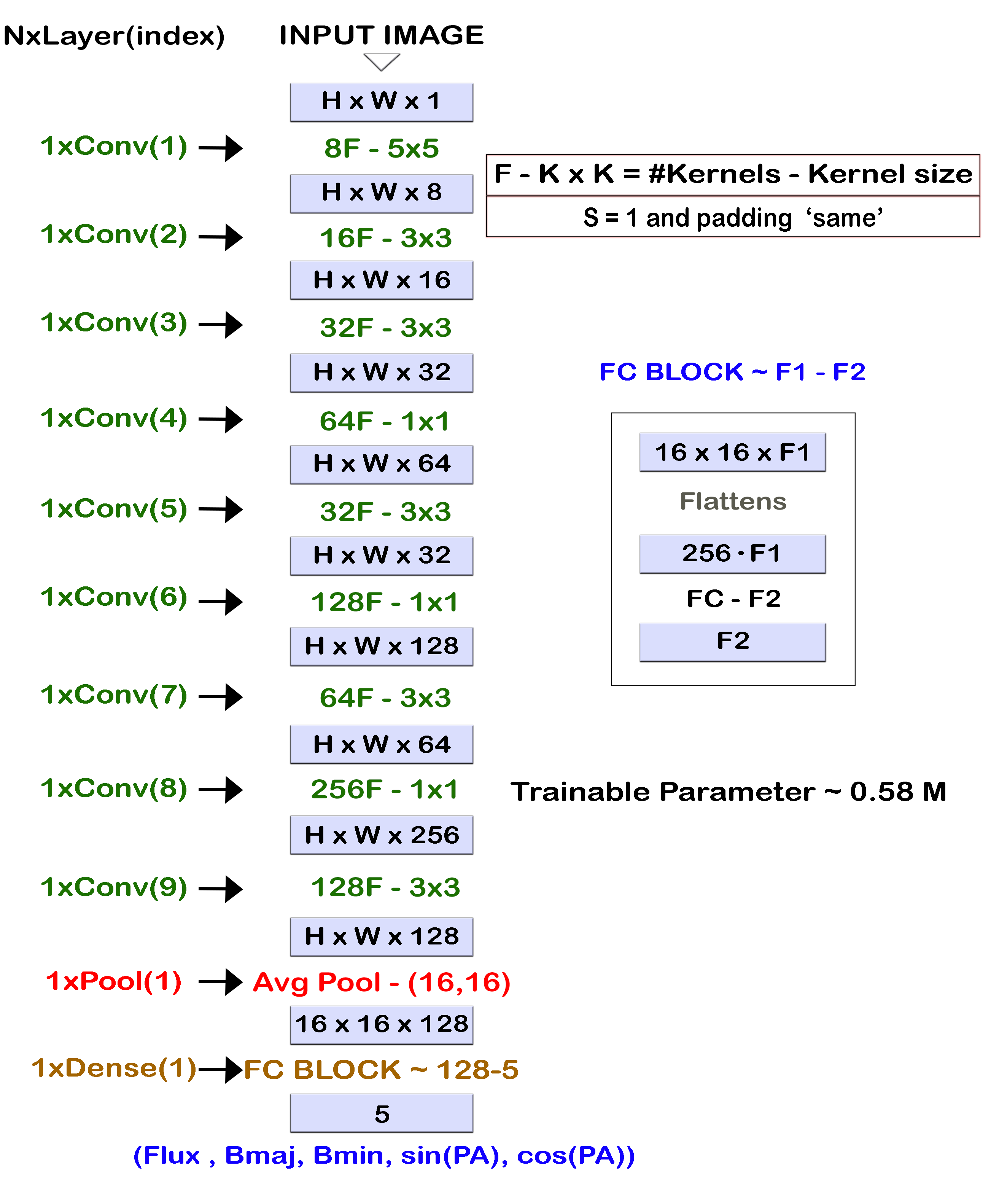}
    \caption{Architecture of the parameter regression model, containing approximately 0.58 million trainable parameters. The network takes a single-channel image and predicts five continuous source parameters: $f_{\rm appr}$, $b_{\rm maj}$, $b_{\rm min}$, $\sin(\phi)$, and $\cos(\phi)$.}
    \label{fig:parameter_model}
\end{figure}

The \texttt{MSC} model predicts five key source parameters: $f_{\rm appr}$, $b_{\rm maj}$, $b_{\rm min}$, $\sin{(\phi)}$, and $\cos{(\phi)}$ from an input image of size $H_{\rm in} \times W_{\rm in} \times 1$. All parameters are normalised to the range $[0,1]$. The overall architecture is shown in Fig.~\ref{fig:parameter_model}. We divide the MSC architecture into initial layers (Conv(1)-(3)), followed by alternating layers (Conv(4)-(9)) with varying kernel sizes, and ending with adaptive average pooling layers. Ultimately, a dense layer containing an FC block extracts the five output parameters. We sequentially discuss the MSC architecture in the following points.
\begin{itemize}
    \item \textbf{Initial layers (Conv(1)–Conv(3)):} Conv(1) uses a $5 \times 5$ kernel size with 8 kernels to capture the large structures of radio sources. A large number of $5\times5$ kernels will increase computational cost and result in slow training. We find that the dataset contains only a few large radio sources, and 8 kernels are sufficient to capture large source features. To capture small features of the sources, we adopt $3 \times 3$ kernels with 16 and 32 kernels in Conv(2) and Conv(3), respectively.
    
    \item \textbf{Alternating layers: Conv(4)–Conv(9):} These layers use alternating $1 \times 1$ and $3 \times 3$ convolutions to refine the feature maps. The $1 \times 1$ convolutions (Conv(4), Conv(6), Conv(8)) with 64, 128, and 256 kernels increase the channel depth while preserving spatial dimensions. This enables deeper combinations of features without significantly increasing computational costs. Larger kernels such as $3 \times 3$ or $5 \times 5$ would be more computationally intensive and slow down the training.
    
    \item \textbf{Pooling layer (Pool1):} An adaptive average pooling layer produces a fixed output of $16 \times 16$, regardless of the input size. It reduces spatial resolution by averaging values within each pooling window, resulting in a feature map of size $16 \times 16 \times 64$. For example, a $128 \times 128$ input uses an $8 \times 8$ window, while a $16 \times 16$ input uses a $1 \times 1$ window.
    
\item \textbf{Dense layer: Dense(1):} The output of Pool(1) is fed into Dense (1) and flattened into a vector of $256 \times 128 = 32768$ features and fed into a fully connected layer (FC) with 5 output neurones representing 5 target parameters. 
\end{itemize}

All CNN layers in the \texttt{MSC} network use the \({ReLU}\) activation function. A small dropout rate of $0.0001\%$ is applied after Conv(3), Conv(5), and Conv(7) to reduce over-fitting by randomly disabling some neurones during training. This helps the network learn more diverse and robust feature representations for better generalisation. The fully connected (FC) layer does not use any activation function, allowing for a direct mapping from the input features to the output parameters. In general, the network contains approximately 0.58 million trainable parameters, making it a lightweight and efficient model suitable for use in resource-constrained environments, with faster training and inference times.

\section{Loss function and training Setup} \label{Loss_function_Setup}
For both detection models, a custom loss function is defined that combines three main components: (1) the bbox regression loss (${L}^{\text{bbox}}$), (2) the source confidence loss (${L}^{\text{obj}}$), and  (3) the non-source or background loss (${L}^{\text{noobj}}$). Since our models do not perform source classification, the classification loss term is excluded. The total loss for the source detection task is expressed as:
\begin{align}
\mathcal{L} &= \sum_{i=0}^{G^2 - 1} \sum_{j=0}^{B - 1} \bigg[ \mathcal{L}^{\text{bbox}}_{i,j} + \mathcal{L}^{\text{obj}}_{i,j} + \mathcal{L}^{\text{noobj}}_{i,j} \bigg].
\label{eq:total_loss}
\end{align}
This equation represents the combined loss of all cells in the grid scale ($G\times G = G^2$) and the anchor boxes ($A$). The loss function \(\mathcal{L}\) is calculated for each pair \((i, j)\). For each anchor, the model predicts five parameters: \((\hat{c}'_{conf}, \hat{x'_c}, \hat{y'_c}, \hat{w'}, \hat{h'})\). These are the raw outputs of the network, which can take any value in the range \((-\infty, +\infty)\). To convert them into meaningful predictions, non-linear exponential mappings are applied. First, a sigmoid activation $\sigma$ is applied to $\hat{c}'_{\rm conf}$ to obtain the confidence score $C'_{\rm conf}$:
\begin{equation}
C'_{\rm conf} = \sigma(\hat{c}'_{\rm conf}),
\label{eq:conf_sigmoid}
\end{equation}
where
\begin{equation}
\sigma(x) = \frac{1}{1 + e^{-x}}.
\label{eq:sigmoid}
\end{equation}
This maps the output to $[0, 1]$, making it suitable for binary classification between the source and background. The confidence score $C'_{\rm conf}$ is used in the focal loss to compute the source loss ($\mathcal{L}^{\rm obj}$), the non-source loss ($\mathcal{L}^{\rm noobj}$) and the box regression loss. In order to convert the raw output \((\hat{x}', \hat{y}', \hat{w}', \hat{h}')\) into the prediction format \(({x'_c}, {y'_c}, {w'}, {h'})\) we first estimate the position with respect to the grid cell using the $\sigma$ to \(\hat{x}', \hat{y}'\) and constrain them within the grid cell. Then the corresponding grid cell coordinates $(g_x, g_y)$ were added to give the absolute position of the source as follows:
\begin{align}
{x'_c}, {y'_c} = (\sigma(\hat{x}') + g_x), (\sigma(\hat{y}) + g_y)
\label{eq:center_coords}
\end{align}
Here, \((g_x, g_y)\) denote the coordinates of the grid cells along the x and y axes with respect to $G$. For \({w}'\) and \({h}'\), we apply an exponential function and scale the result by the corresponding anchor box width ($a_w$) and height ($a_h$). This provides $w'$ and $h'$ in relation to the encoded ground truth ($w$, $h$) as described in Section~\ref{Data_encoding}.
\begin{equation}
w' = a_w \exp(\hat{w}),
\qquad
h' = a_h \exp(\hat{h}),
\label{eq:width_height}
\end{equation}

In the next three subsections, we discuss the loss considered over bboxes, source and background losses, and parameter prediction. The fourth and fifth subsections address training setups for \texttt{SSD}, \texttt{MSD}, and \texttt{MSC} models.

\subsection{Complete Intersection of Union Loss (CIoU)}
The final prediction parameters $({C'_{\rm conf}}, {x'_c}, {y'_c}, {h'}, {w'})$ for each anchor are compared with their corresponding encoded gt-box ($C_{\rm conf}$, $x_{\rm c}$, $y_{\rm c}$, $h$, $w$) to calculate the loss using the Complete Intersection over Union (CIoU) loss~\citep{zheng2019diou}, as defined in Eq.~\eqref{eq:ciou_loss}. CIoU loss is an advanced bounding box regression metric designed to overcome the limitations of traditional IoU-based losses. Although the standard IoU measures the overlap between the predicted and gt-bboxes, it fails to provide useful gradient information when there is no overlap. CIoU addresses this issue by incorporating additional geometric factors, making it more effective and stable during training. Specifically, CIoU loss considers four key aspects: the overlap area, the distance between the centres of the gt-bbox and the predicted box, the diagonal length of the enclosing box, and the consistency of their aspect ratios. This makes it particularly useful for sparse-object detection tasks, such as those in astronomical images, where precise localisation is essential.

Let $i \in \{0,\dots,G^2-1\}$ index the grid cells and $j \in \{0,\dots,B-1\}$ index the anchors. The gt-bbox associated with the grid cell $i$ and the anchor $j$ is
\[
\mathbf{B}_{i,j} = \big(x_{c,(i,j)},\, y_{c,(i,j)},\, w_{i,j},\, h_{i,j}\big),
\]
and the corresponding predicted box is
\[
\mathbf{B}'_{i,j} = \big(x'_{c,(i,j)},\, y'_{c,(i,j)},\, w'_{i,j},\, h'_{i,j}\big).
\]

The CIoU loss is defined as
\begin{align}
\mathcal{L}^{\mathrm{bbox}}_{i,j}
= \mathbf{1}^{\mathrm{obj}}_{i,j} \Bigg[ 
& 1 - \mathrm{IoU}(\mathbf{B}_{i,j}, \mathbf{B}'_{i,j}) \nonumber \\
& + \frac{\rho^2\big((x_{c,(i,j)}, y_{c,(i,j)}), (x'_{c,(i,j)}, y'_{c,(i,j)})\big)}{l^2} \nonumber \\
& + \alpha v 
\Bigg].
\label{eq:ciou_loss}
\end{align}
where $\mathbf{1}^{\mathrm{obj}}_{i,j}$ indicates whether anchor $j$ in the grid cell $i$ is assigned to a source:
\begin{equation}
\mathbf{1}^{\mathrm{obj}}_{i,j} =
\begin{cases}
1, & \text{source present}, \\
0, & \text{otherwise}.
\end{cases}
\label{eq:indicator_obj}
\end{equation}

The squared Euclidean distance between the box centres is
\begin{equation}
\rho^2
= (x_{c,(i,j)} - x'_{c,(i,j)})^2
+ (y_{c,(i,j)} - y'_{c,(i,j)})^2,
\label{eq:rho_squared}
\end{equation}
and $l^2$ denotes the squared diagonal length of the smallest enclosed box that covers both $\mathbf{B}_{i,j}$ and $\mathbf{B}'_{i,j}$. The weighting factor $\alpha$ is given by
\begin{equation}
\alpha = \frac{v}{(1 - \mathrm{IoU}) + v},
\label{eq:alpha_ciou}
\end{equation}
where
\begin{equation}
v = \frac{4}{\pi^2}
\left[
\arctan\!\left(\frac{w'_{i,j}}{h'_{i,j}}\right)
- \arctan\!\left(\frac{w_{i,j}}{h_{i,j}}\right)
\right]^2.
\label{eq:v_ciou}
\end{equation}
This term penalises shape mismatch, encouraging improved geometric consistency even at high overlap.
\subsection{Focal loss for source and background}

We adopt focal loss \citep{Focal_loss} to address the imbalance between the pair of grid anchors assigned to the positive anchors (sources) and those corresponding to the negative anchors (background). Let $C'_{\rm conf,(i,j)} \in (0,1)$ denote the predicted confidence for anchor $j$ in grid cell $i$, and let $C_{\rm conf,(i,j)} \in \{0,1\}$ be the corresponding ground-truth label, where $C_{\rm conf,(i,j)}=1$ indicates a source.

The focal loss for the source is
\begin{align}
\mathcal{L}^{\mathrm{obj}}_{i,j}
= -\,\beta_{i,j}\,\mathbf{1}^{\mathrm{obj}}_{i,j}\,
\alpha' \,(1 - C'_{\rm conf,(i,j)})^{\gamma}
\log\!\big(C'_{\rm conf,(i,j)}\big),
\label{eq:conf_obj}
\end{align}
and for background anchors;
\begin{align}
\mathcal{L}^{\mathrm{noobj}}_{i,j}
= -\,\beta_{i,j}\,\mathbf{1}^{\mathrm{noobj}}_{i,j}\,
(1 - \alpha')\,(C'_{\rm conf,(i,j)})^{\gamma}
\log\!\big(1 - C'_{\rm conf,(i,j)}\big).
\label{eq:conf_noobj}
\end{align}
The imbalance factor $\alpha' \in [0,1]$ controls the relative weighting of the source and background anchors, with larger values emphasising sources. We adopt $\alpha' = 0.75$ for both SSD and MSD models to prioritise source-related focal loss. The focusing parameter is fixed to $\gamma = 2$, which suppresses the loss contribution from well-classified source and background anchors and emphasises harder, misclassified examples. The modulation terms $(1-C'_{\rm conf,(i,j)})^\gamma$ and $(C'_{\rm conf,(i,j)})^\gamma$ suppress the loss from well-classified source and background anchors, respectively. The additional weighting factor
\[
\beta_{i,j} = \big(C_{\rm conf,(i,j)} - C'_{\rm conf,(i,j)}\big)^2
\]
further scales the loss, assigning larger weights to anchors with higher prediction errors.

The indicator for non-object anchors is defined as
\begin{align}
\mathbf{1}^{\mathrm{noobj}}_{i,j}
= \big(1 - \mathbf{1}^{\mathrm{obj}}_{i,j}\big)\,
\mathbf{1}\!\left[\mathrm{DIoU}_{i,j} < \mathrm{DIoU}_{\mathrm{thresh}}\right],
\label{eq:indicator_noobj}
\end{align}
where $\mathbf{1}^{\mathrm{noobj}}_{i,j}$ is active only for anchors not assigned to any source and with a Distance-IoU below the threshold $\mathrm{DIoU}_{\mathrm{thresh}} = 0.5$.

The Distance-IoU (DIoU) metric is defined as
\begin{align}
\mathrm{DIoU}
= \mathrm{IoU} - \frac{\rho^2}{l^2},
\label{eq:diou}
\end{align}
where $\rho^2$ is the Euclidean distance squared between the centre of the gt-bbox and the predicted bbox. $l^2$ is the diagonal square length of the smallest enclosed box. This formulation penalises spatially distant predictions even when overlap exists, thereby improving localisation accuracy while increasing confidence for true sources and suppressing background responses. The final detection loss is obtained by combining the equations~\eqref{eq:ciou_loss}, \eqref{eq:conf_obj}, and \eqref{eq:conf_noobj} into the total loss defined in equation~\eqref{eq:total_loss}, which is minimised during training to jointly optimise localisation and confidence prediction. During inference, the encoded predictions are rescaled to the original $512 \times 512$ image resolution by multiplying $\Delta$ of the corresponding $G$, as described in Section~\ref{Data_encoding}.

\subsection{Mean Squared Error Loss}
For the MSC model, we adopt the mean squared error (MSE) loss to regress five continuous source parameters: \(f_{\rm appr}\), \(b_{\rm maj}\), \(b_{\rm min}\), \(\sin\phi\), and \(\cos\phi\). For a single training sample, the loss is defined as
\begin{align}
\mathcal{L}_{\mathrm{char}} =
\frac{1}{5} \sum_{k=1}^{5} \left(y'_k - y_k\right)^2,
\label{eq:char_loss}
\end{align}
where \(y'_k\) and \(y_k\) denote the predicted and ground-truth values of the \(k\)-th parameter, respectively. This formulation assigns equal weight to the regression error of each parameter.

\subsection{Training and Validation of Detectors}
\begin{figure}
    \centering
    \includegraphics[width=1.\linewidth]{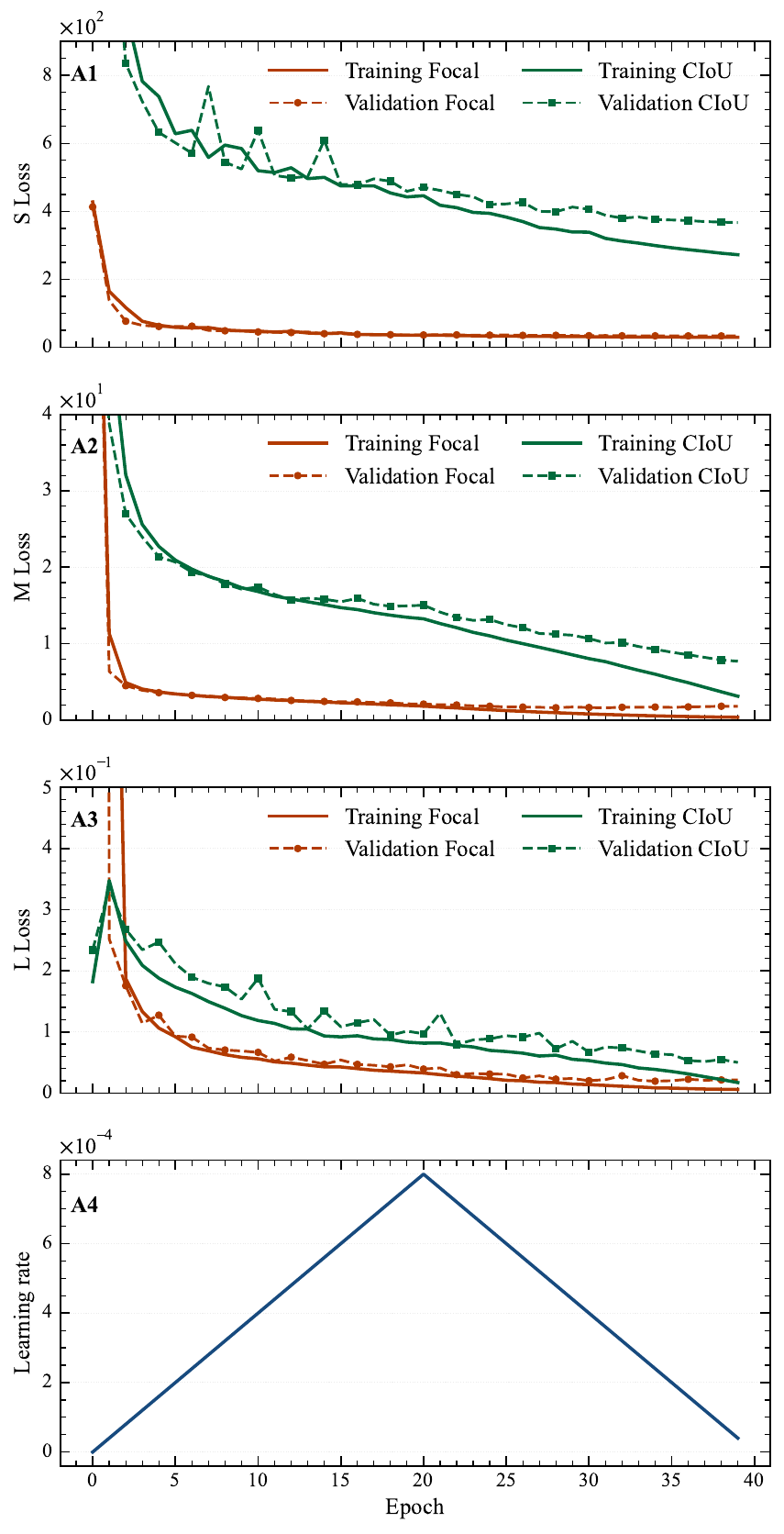}
    \caption{Panels A1--A3 show the training (solid lines) and validation (dashed lines) curves for CIoU Loss (green) and Focal Loss (orange) corresponding to S Loss, M Loss, and L Loss, respectively. The epoch axis ranges from 0 to 39, encompassing a total of 40 epochs. Panel A4 presents the corresponding learning-rate schedule.}
    \label{fig:loss_Small}
\end{figure} 
We train both detectors for 40 epochs with the Adam optimiser \citep{kingma2017adam} with default parameters ($\beta_1 = 0.9$, $\beta_2 = 0.999$, $\epsilon = 10^{-8}$) and a cyclical learning rate ($\eta_t$) \citep{smith_2017_CLR} defined as:
\begin{equation}
\eta_t = \eta_{\text{base}} + (\eta_{\text{max}} - \eta_{\text{base}}) \cdot \max(0, 1 - |x|),
\label{eq:Cyclic_LR}
\end{equation}
where $x = \frac{t}{s} - 2 \cdot \left\lfloor \frac{t}{2s} \right\rfloor + 1$. We define $t$ as the epoch index (starting from $0$), and $s$ is the step size. A complete cycle consists of \(2s\) epochs. The $\eta_t$ increases linearly from \(\eta_{\text{base}} = 10^{-8}\) to \(\eta_{\text{max}} = 8\times10^{-4}\) during the first half \(s = 20\) epochs, then decreases back to \(\eta_{\text{base}}\) during the next \(s\) epochs, forming a triangular cycle (Fig. \ref{fig:loss_Small} A4). This pattern repeats every 40 epochs, promoting better convergence and generalisation by allowing the optimiser to explore wider regions of the loss functional space. Training and validation losses for the CIoU and Focal components, along with the CLR schedule, are shown in Figure~\ref{fig:loss_Small}. Here, panel A1 shows the loss corresponding to the SSD, defined as loss S, while panels A2 and A3 show the loss corresponding to the MSD model for Medium (M) and Large (L) sources, respectively. Training loss (solid lines) and validation loss (dashed lines) for focal loss and CIoU are also shown. The model weights were saved after each epoch. The SSD model was trained on an NVIDIA L4 GPU, requiring approximately \textbf{1.2 hours} for 40 epochs, whereas the MSD model was trained on an NVIDIA Tesla T4 Tensor Core GPU, requiring approximately \textbf{1 hour} for 40 epochs.

\subsection{Training and Validation of the Characterisation model}
\begin{figure}
    \centering
    \includegraphics[width=1.0\linewidth]{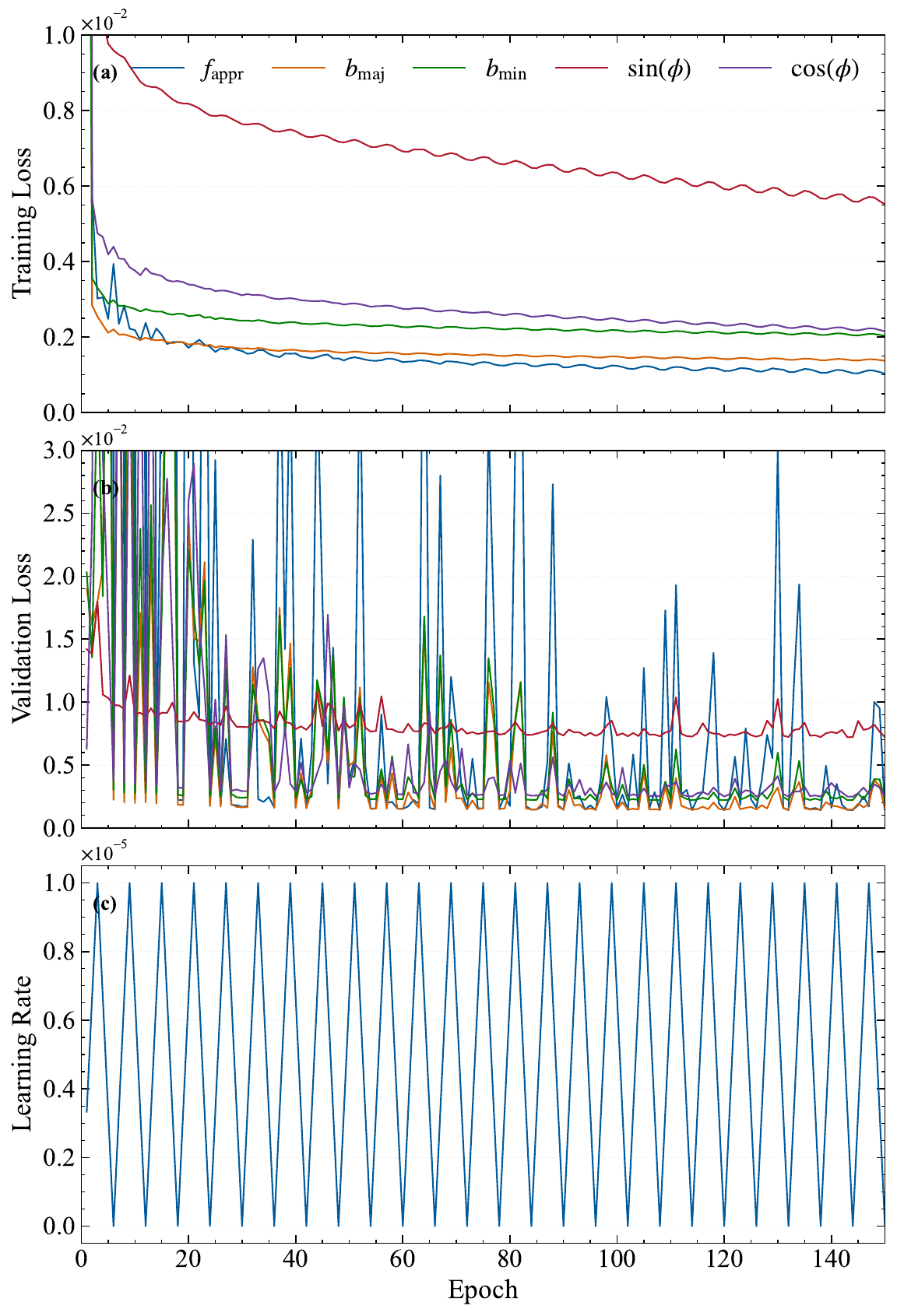}
    \caption{Component-wise loss curves for the five predicted parameters: \(f_{\rm appr}\), \(b_{\rm maj}\), \(b_{\rm min}\), \(\sin(\phi)\), and \(\cos(\phi)\). The top and middle panels display the training and validation losses, respectively, while the bottom panel shows the learning rate schedule. The training losses are plotted over the range [0,\,0.010] and the validation losses over [0,\,0.030]. The model is trained for 150 epochs. This visualisation highlights overall trends and differences in convergence behaviour across components.}
    \label{fig:train_val_loss}
\end{figure}
For the MSC model, we used training samples with redefined sizes, as shown in Table~\ref{tab:uniform_w_h}, together with the number of examples. For the validation, we use the same training region but with non-overlapping regions. In total, we train the \text{MSC} model with 98028 training samples and 26,021 validation samples with a batch size of 16. The model was trained using the cyclic learning rate schedule defined in Equation~\eqref{eq:Cyclic_LR} along with the Adam optimiser using the default parameters, but with \(\eta_{\text{base}} = 10^{-8}\), \(\eta_{\text{max}} = 1\times10^{-5}\), and a smaller step size \(s = 3\). This adaptive learning rate allowed for smooth and stable convergence across all sizes. Training and validation loss with the $\eta_t$ curve is shown in Figure~\ref{fig:train_val_loss}. It provides an overview of the model's performance in epochs, as well as the training and validation losses for different physical parameters. Among these, the \(\sin(\phi)\) component exhibits the highest loss in both training and validation, followed by \(\cos(\phi)\). The spikes and fluctuations observed in \(\sin(\phi)\), \(\cos(\phi)\), and \(f_{\rm appr}\) are likely a consequence of the scaling applied during training: angle-related losses \(\sin(\phi)\), \(\cos(\phi)\) were down weighted by 0.5 such that the model should also focus on the other parameter losses. The loss \(f_{\rm appr}\) was heavily weighted by $10$ to improve accuracy. Additionally, during training data generation, \(\phi\) is set to $0$ for sources with \( b_{\rm maj}\leq1\arcsec \). This choice reduced training noise for these unresolved sources, allowing the model to focus more effectively on learning other physical parameters, such as \(f_{\rm appr}\), \(b_{\rm maj}\), and \(b_{\rm min}\), to improve the overall characterisation performance. Training was performed on an NVIDIA Tesla T4 Tensor Core GPU. The MSC model required approximately \textbf{1 hour} for 150 epochs.

\section{Analysis and Results} \label{Analysis_and_Results}
\begin{figure}
    \centering
    \includegraphics[width=1.00\linewidth]{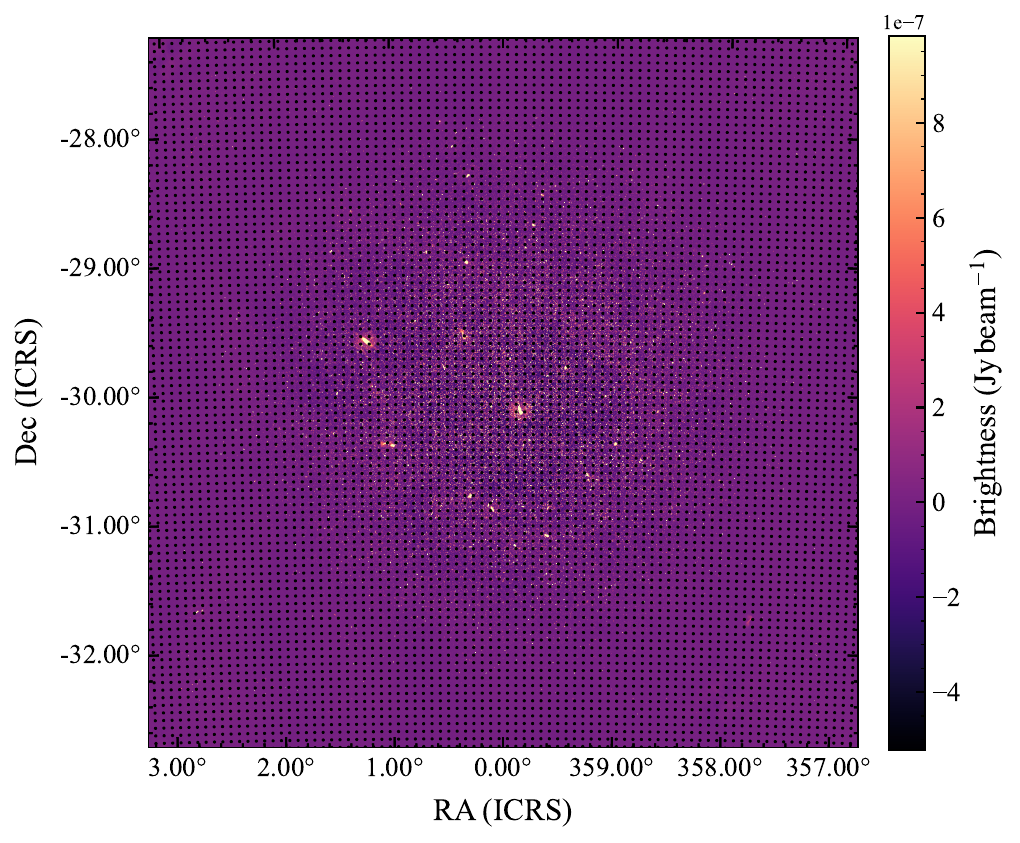}
    \caption{Dense grid in RA and Dec space. RA ranges from \(-3.2^\circ\) to \(3.2^\circ\) with 96 equally spaced points, and Dec ranges from \(-32.8^\circ\) to \(-27.2^\circ\) with 84 equally spaced points. This results in 8,232 total grid points, each used to extract a \(512 \times 512\) image patch for the final generalised epoch of all three models.}
    \label{fig:with_large_grid_points}
\end{figure}

\begin{table*}
\centering
\caption{Comparison of detection performance across optimal epochs, combined multi-scale configurations, and results from other teams on the SKA SDC1 dataset. Purity is defined as the fraction of matched detections among total detections (see Sect.~\ref{Scoring_procedure} for full metric definitions). The best configuration from our framework is highlighted in bold.}
\vspace{1em}
\begin{tabular*}{\textwidth}{@{\extracolsep{\fill}} l r r r r r r c @{}}
\toprule
\textbf{Model / Team / Epoch} & \textbf{F\textsubscript{sco}} & \textbf{N\textsubscript{det}} & \textbf{N\textsubscript{match}} & \textbf{N\textsubscript{false}} & \textbf{N\textsubscript{bad}} & \textbf{Purity} & \textbf{$\bar{w}$} \\
\midrule
\multicolumn{8}{c}{\textbf{Section A: SSD Standalone (\(C'_{\rm conf}\)} $\geq$ 0.5)} \\
\midrule
Epoch 38 & 478634 & 669858 & 631021 & 38837 & 6890 & 94.20\% & 0.8200 \\
\midrule
\multicolumn{8}{c}{\textbf{Section B: Combined Multi-scale Detection (MSC fixed at epoch 92)}} \\
\midrule
Single-Small (0.5) & 458967 & 646644 & 607765 & \textbf{38879} & 6861 & 93.99 & 0.8190 \\
Multiscale-Medium (0.4) & 25866 & 42778 & 38147 & 4631 & 1373 & 89.17 & 0.7995 \\
Multiscale-Large (0.3) & 304 & 574 & 520 & 54 & 17 & 90 & 0.6884 \\
All-Combined Result & \textbf{483875} & \textbf{689996} & \textbf{645650} & \textbf{44346} & \textbf{8148} & \textbf{93.57} & \textbf{0.8181} \\
\midrule
\multicolumn{8}{c}{\textbf{Section C: Other Team Results}} \\
\midrule
MINERVA (YOLO-CIANNA) & 480450 & 724480 & 680000 & 44480 & 16839 & 93.86\% & 0.7719 \\
MINERVA (purity-based) & 418434 & 541542 & 536412 & 5130 & 2506 & 99.06\% & 0.7896 \\
JLRAT2 (JSFM2) & 298201 & 502146 & 484212 & 17934 & 2274 & 96.43\% & 0.6529 \\
Engage-SKA (PROFOUND) & 200939 & 421992 & 418384 & 3608 & 2677 & 99.15\% & 0.4889 \\
Shanghai (Multiple Methods) & 158841 & 292646 & 291553 & 1093 & 698 & 99.63\% & 0.5486 \\
ICRAR (CLARAN) & 142784 & 279898 & 259806 & 20092 & 6875 & 92.82\% & 0.6269 \\
\textbf{Ours (Best Config)} & \textbf{483875} & \textbf{689996} & \textbf{645650} & \textbf{44346} & \textbf{8148} & 93.57
 & \textbf{0.8181}\\
\midrule
\multicolumn{8}{c}{\textbf{Section D: combined catalogue with varying $C'_{\rm conf}$}} \\
\midrule
{$C'_{\mathrm{conf}} \geq 0.5$} & 484120 & 680925 & 640500 & 40425 & 7296 & 94.06 & 0.8190 \\
{$C'_{\mathrm{conf}} \geq 0.55$} & 458845 & 590410 & 574920 & 15490 & 3224 & 97.38 & 0.8250 \\
{$C'_{\mathrm{conf}} \geq 0.6$} & 412431 & 508771 & 502982 & 5789 & 1265 & 98.86 & 0.8315 \\
{$C'_{\mathrm{conf}} \geq 0.65$} & \textbf{347236} & \textbf{418291} & \textbf{416326} & \textbf{1965} & \textbf{447} & \textbf{99.53} & \textbf{0.8388} \\
{$C'_{\mathrm{conf}} \geq 0.7$} & 255165 & 302252 & 301732 & 520 & 120 & 99.83 & 0.8473 \\
{$C'_{\mathrm{conf}} \geq 0.75$} & 129506 & 151172 & 151058 & 114 & 28 & 99.92 & 0.8581 \\
\bottomrule
\end{tabular*}
\label{tab:combined_results_equal_spacing}
\end{table*}

\begin{table*}
\centering
\caption{Sensitivity of the catalogue-merging procedure to the size of the centred square boxes used during the final NMS stage with an IoU threshold of 0.1. The catalogue was constructed using the best detector configuration ($C'_{\mathrm{conf}}\geq0.5$ for S-boxes, $C'_{\mathrm{conf}}\geq0.4$ for M-boxes, and $C'_{\mathrm{conf}}\geq0.3$ for L-boxes). Smaller square boxes result in insufficient merging of duplicate detections, whereas larger square boxes increase the merging of nearby detections. The $4\times4$ pixel square boxes provide the highest overall SKA SDC1 score, $F_{\rm sco}$, and were therefore adopted throughout this work.}
\label{tab:merge_box_size}

\begin{tabular*}{\textwidth}{@{\extracolsep{\fill}} l r r r r r r c @{}}
\hline
\textbf{Square box} &
$\mathbf{F_{\rm sco}}$ &
$\mathbf{N_{\rm det}}$ &
$\mathbf{N_{\rm match}}$ &
$\mathbf{N_{\rm false}}$ &
$\mathbf{N_{\rm bad}}$ &
\textbf{Purity (\%)} &
\textbf{$\bar{w}$} \\
\hline
$2\times2$ & 473123 & 706263 & 648699 & 57564 & 8456 & 91.84 & 0.8180 \\
\textbf{$4\times4$} & \textbf{483875} & 689996 & 645650 & 44346 & 8148 & 93.57 & 0.8181 \\
$6\times6$ & 482036 & 684179 & 641337 & 42842 & 8005 & 93.73 & 0.8184 \\
$8\times8$ & 471055 & 661131 & 622292 & 38839 & 7443 & 94.12 & 0.8193 \\
\hline
\end{tabular*}
\end{table*}

\begin{table*}
\centering
\caption{Characterisation performance as a function of source compactness and SNR. For matched sources, the table reports the median fractional errors in $f$, $b_{\rm maj}$, and $b_{\rm min}$ (expressed as percentages), together with the mean absolute error and standard deviation of the $\phi$ residuals for the best-performing \texttt{YOLO-CHARS} catalogue (catalogue purity of 93.57\%). Corresponding residual distributions for the $2.738^\circ \times 2.35^\circ$ benchmark field, using a \texttt{YOLO-CHARS} catalogue selected with $C'_{\rm conf}\geq0.64$ to achieve a catalogue purity comparable to that of PyBDSF, are presented in Fig.~\ref{fig:catalog_comparison_histograms} of Appendix~\ref{appendix_3}.}
\label{tab:characterisation_results}

\begin{tabular*}{\textwidth}{@{\extracolsep{\fill}} l c c c c c c @{}}
\hline
Population &
$N_{\rm match}$ &
Median $\left|\frac{\Delta f}{f_{\rm true}}\right|$ (\%) &
Median $\left|\frac{\Delta b_{\rm maj}}{b_{\rm maj,true}}\right|$ (\%) &
Median $\left|\frac{\Delta b_{\rm min}}{b_{\rm min,true}}\right|$ (\%) &
$\langle |\Delta\phi| \rangle$ &
$\sigma_{\Delta\phi}$\\
&&&&&
(\degree) &
(\degree) \\
\hline
Compact ($5$--$8\sigma$)     & 23\,670  & 33.15 & 56.61 & 48.66 & 45.15 & 52.09 \\
Extended ($5$--$8\sigma$)    & 42       & 40.58 & 90.39 & 94.31 & 43.87 & 51.65 \\
Compact ($8$--$15\sigma$)    & 98\,040  & 16.50 & 39.54 & 41.49 & 44.79 & 51.79 \\
Extended ($8$--$15\sigma$)   & 344      & 27.17 & 80.41 & 84.36 & 44.86 & 52.14 \\
Compact ($>15\sigma$)        & 420\,454 & 11.63 & 29.29 & 35.51 & 40.10 & 48.02 \\
Extended ($>15\sigma$)       & 98\,460  & 19.27 & 28.16 & 40.05 & 17.61 & 26.86 \\
\hline
\end{tabular*}
\end{table*}

Using the epoch-analysis procedure described in the Appendix~\ref{epoch_analysis}, we determined the optimal training epoch for each model based on its performance on a subfield. All models were optimised with respect to \(F_{\rm sco}\), purity, and \(\bar{w}\). From this analysis, the optimal epochs for the \texttt{SSD (S-boxes)}, \texttt{MSD (M-boxes)} and \texttt{MSD (L-boxes)} models were found to be 38, 33 and 34, respectively, while the \texttt{MSC} model achieved its best convergence at epoch~92. 

Then we use the following step to build the SSD and MSD catalogues and combine them:

\textbf{Step 1}: To ensure complete and uniform coverage, we create a grid of 84 equally spaced points along the RA axis and 96 along the Dec axis, spanning from $-3.2^{\circ}$ to $3.2^{\circ}$ in RA and from $-32.8^{\circ}$ to $-27.2^{\circ}$ in Dec (see Fig.~\ref{fig:with_large_grid_points}). For each grid point, we extract a square cutout of $309.5\arcsec \times 309.5\arcsec$ (512 $\times$ 512 pixels) centred on that RA and Dec coordinate. Cutouts lying at the edges or outside the full map region are discarded. This process generates approximately 1008 batches with a batch size of 8. The cutouts are highly overlapping, allowing the model to generate predictions for every region of the image. \textbf{Step 2}: We first select the SSD model independently and load the weights from the optimal epoch. Predictions are performed batch-wise (batch size 8) using a confidence threshold $C'_{\mathrm{conf}} \geq 0.5$. Since multiple detections may correspond to the same source, NMS is applied with an IoU threshold of 0.2, keeping only the detection with the highest $C'_{\mathrm{conf}}$ for each source. \textbf{Step 3}: After obtaining all detections from the highly overlapping patches in Step 2, we build the SSD catalogue by converting each source's local centre pixel coordinates to global sky coordinates (RA, Dec) and recording the bounding box width \(W'\) and height \(H'\). Because the overlapping patches often produce duplicate detections of the same source, the initial catalogue contains many duplicates. To remove them, NMS is again applied with an IoU threshold of 0.2. This produces a clean, non-duplicate set of SSD detections. \textbf{Step 4}: Similarly, the MSD model was run on the same image cutouts as in Step~2, using the same batch size. Medium-sized sources were detected from output $O_{14}$ (M-boxes) at its optimal epoch of 33, while large-sized sources were detected from output $O_{20}$ (L-boxes) at the optimal epoch of 34. These outputs generate 9216 and 2304 detections per image, respectively (see Fig.~\ref{fig:model_architecture_M}). \(C'_{\rm conf}\) was set to \(\geq0.4\) for M-boxes and \(\geq0.3\) for L-boxes. Similarly, NMS was applied independently to the M-boxes and L-boxes catalogues following the procedure described in Steps~2 and 3, using an IoU threshold of 0.1 to eliminate multiple and duplicate detections of the same source. \textbf{Step 5}: The final merged catalogue is constructed by combining detections from the SSD (small-scale) and MSD (M-boxes and L-boxes) models obtained in Steps~3 and 4. A source may be detected by SSD and in one or both MSD outputs (M-boxes and/or L-boxes) with different bounding box extents and associated \(C'_{\mathrm{conf}}\) values. Because both detectors are trained independently with different output scales, \(C'_{\mathrm{conf}}\) cannot be directly compared across the detector outputs. Applying NMS directly to the detected bboxes could lead to the rejection of valid lower-confidence detections whenever their IoU exceeds 0.1 with a higher-confidence detection from another branch output, even when the corresponding bboxes have different spatial extents and centre locations. 
To avoid this issue and preserve all meaningful detections, we adopt the following strategy. For every detection in the merged catalogue, irrespective of scale, we generate a centred \(4\times4\)-pixel square box, corresponding to an angular size of \(2.4\arcsec\times2.4\arcsec\), and apply NMS with an IoU threshold of 0.1 using only these boxes. This suppresses multiple detections originating from different detector outputs when they correspond to the same source or have nearly identical centroids, while retaining detections with spatially distinct centres.

To assess the sensitivity of the catalogue-merging procedure to the adopted box size, we repeated the analysis using sizes of $2\times2$, $6\times6$, and $8\times8$ pixels. For the $2\times2$ pixel boxes, the total number of detections ($N_{\rm det}$) increased, but so did the number of $N_{\rm false}$ and $N_{\rm bad}$. In many cases, the boxes were too small to satisfy the adopted IoU threshold of 0.1, resulting in insufficient merging and a lower $F_{\rm sco}$. Increasing the box size to $6\times6$ and $8\times8$ pixels reduced both $N_{\rm det}$ and $F_{\rm sco}$, as larger boxes increased the likelihood of merging nearby detections that should remain distinct. We therefore adopt the $4\times4$ pixel boxes as the optimal compromise between suppressing multiple detections and preserving neighbouring sources. The quantitative results for the different box sizes are presented in Table~\ref{tab:merge_box_size}.

Although the proposed strategy provides robust performance for the SDC1 data, the original predicted bboxes are replaced by centred \(4\times4\) pixel boxes during the final catalogue-merging stage. Consequently, closely separated or physically associated sources may occasionally be merged or retained incorrectly. Its limitations and potential improvements are discussed in Section~\ref{sec:discussion}.

The final resulting catalogue contains one entry per unique source, with fields: $\mathrm{id}$, $C'_{\mathrm{conf}}$, $M_{\mathrm{class}}$ (1 for SSD, 2 for MSD M-boxes, 3 for MSD L-boxes), $\mathrm{RA}$, $\mathrm{Dec}$, $X'_{\mathrm{c}}$, $Y'_{\mathrm{c}}$, $W'$, $H'$. For extended sources detected by MSD, compact sources detected by SSD that fall within the larger bbox are retained as separate entries based on $M_{\mathrm{class}}$ unless suppressed by NMS. During catalogue construction, detections from different detector outputs are merged only when identified as duplicates by the catalogue-merging procedure. The framework, therefore, does not attempt to determine whether nearby radio components are physically associated (e.g. a core and its lobes belonging to the same galaxy) or represent unrelated neighbouring sources projected close together on the sky. All retained detections are preserved as individual catalogue entries, allowing source association to be performed subsequently using additional morphological, positional, or contextual information. Consequently, the present framework is designed for source detection and characterisation rather than component association.

After obtaining the independent SSD catalogue and the combined catalogue, we prepare each detected bbox for input to the MSC model. The bbox is resized to a fixed size $(W_{\mathrm{in}}, H_{\mathrm{in}})$ and normalized using Eq.~\ref{eq:ch_log_norm_unit} (see Sect.~\ref{Char_Data_Preprocessing} for details). A masking step is then applied to exclude contributions from neighbouring sources that fall inside or overlap the target source bbox. This step is essential in crowded regions where compact sources are located close to, or within, larger extended sources, thereby ensuring accurate parameter estimation. The MSC model is the only multi-scale characterisation network trained to predict physical parameters simultaneously across angular scales of \(0.1\arcsec \leq b_{\rm maj} \leq 128\arcsec\).

\begin{figure}
    \centering
    \includegraphics[width=\linewidth]{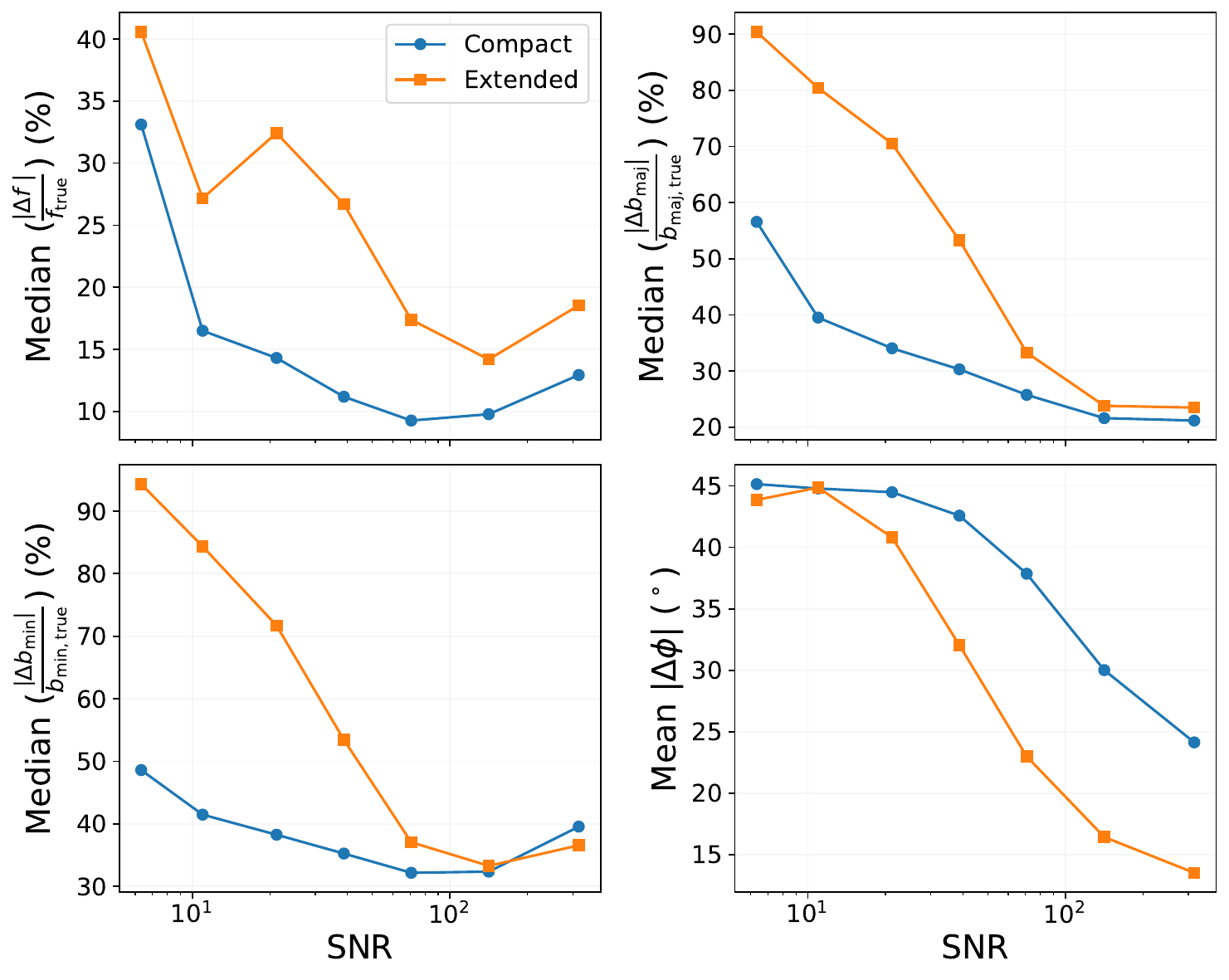}
    \caption{Characterisation performance as a function of SNR for compact and extended sources corresponding to the best-performing \texttt{YOLO-CHARS} catalogue. The panels show the median fractional errors in \(f\), \(b_{\rm maj}\), and \(b_{\rm min}\), together with the mean absolute error (MAE) in \(\phi\), computed from matched sources. The errors generally decrease with increasing SNR. Extended sources exhibit systematically larger fractional errors than compact sources, except for \(\phi\), for which the MAE is larger for compact sources because they are less well resolved.} \label{fig:characterisation_vs_snr}
\end{figure}

\begin{figure}
    \centering
    \includegraphics[width=\linewidth]{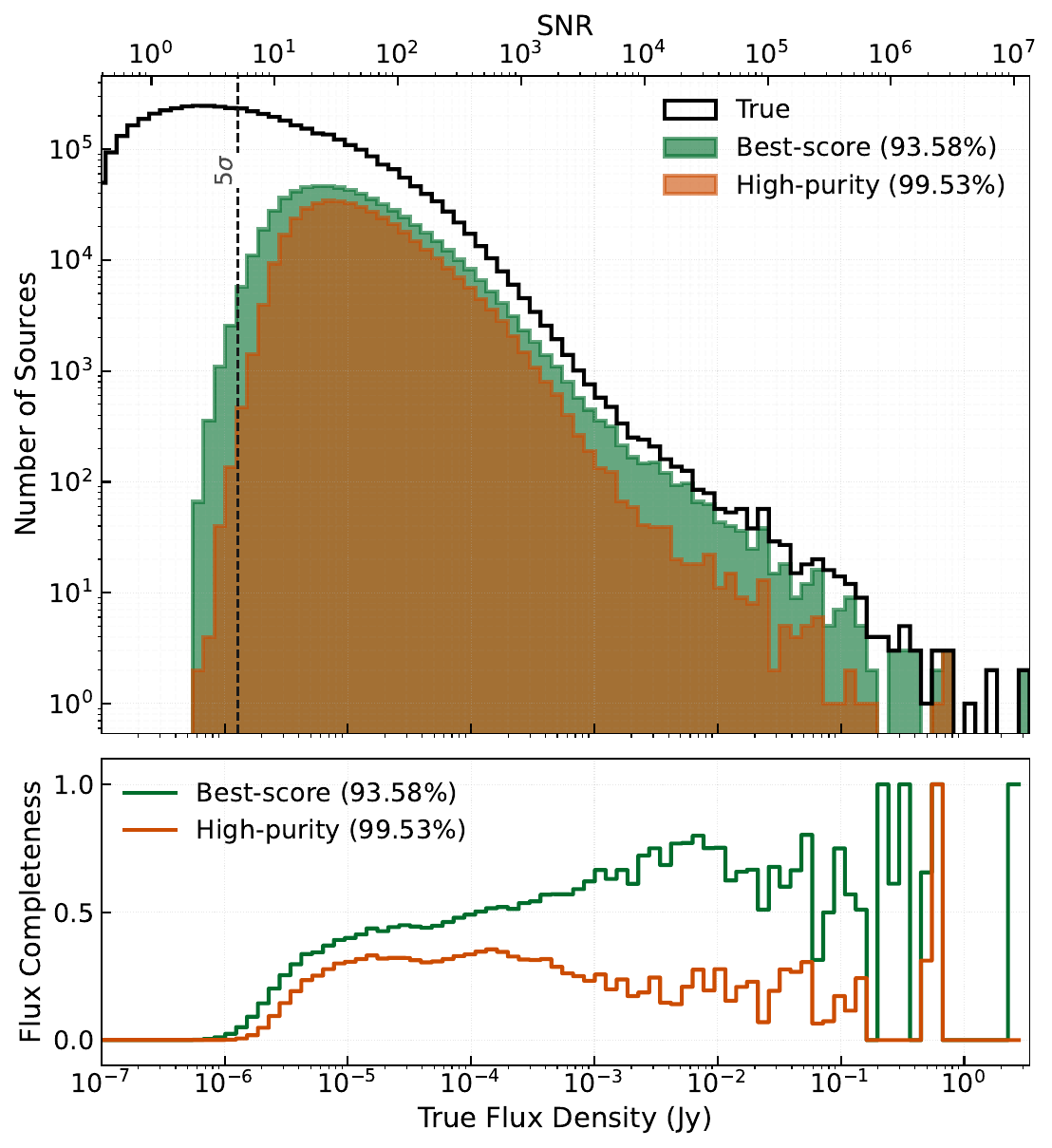} 
    \caption{Source recovery and completeness as a function of true flux density for the best-scoring catalogue (purity = 93.57 per cent; Table~\ref{tab:combined_results_equal_spacing}, Section~C) and the high-purity catalogue ($C'_{\rm conf}\geq0.65$; purity = 99.53 per cent; Table~\ref{tab:combined_results_equal_spacing}, Section~D). The upper panel shows the number of sources in each true-flux-density bin for the reference catalogue and the corresponding number of matched sources recovered by each configuration. The lower panel shows the completeness, defined as the fraction of reference-catalogue sources recovered in each true flux-density bin, i.e. $N_{\rm match}/N_{\rm true}$. The upper x-axis indicates the corresponding SNR.}
    \label{fig:completeness}
\end{figure}

\begin{figure}
    \centering
    \includegraphics[width=1.0\linewidth]{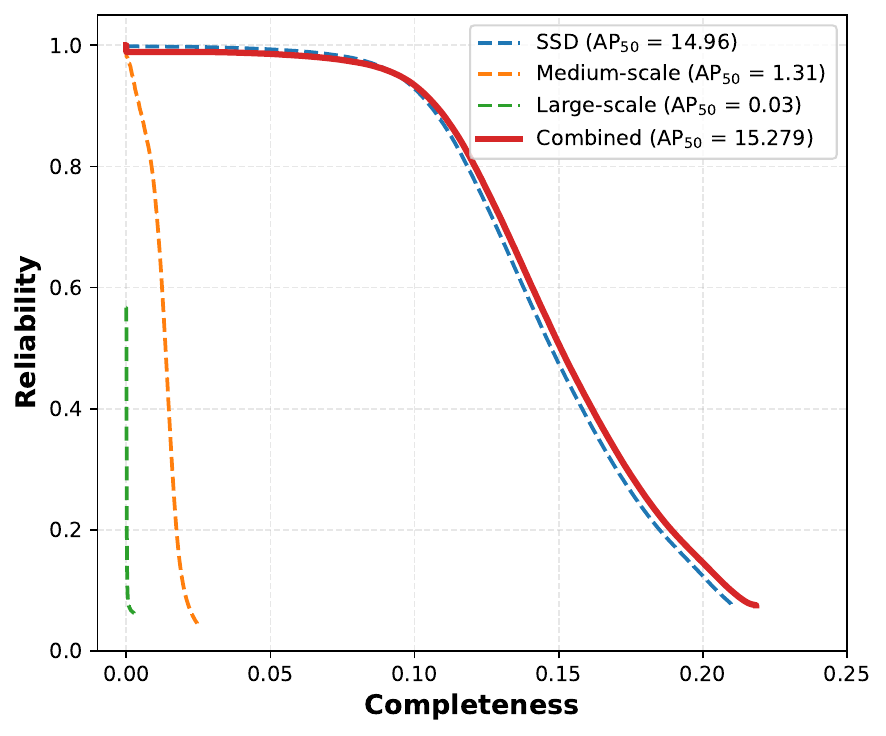}
    \caption{$AP_{50}$ reliability--completeness curves for the SSD model, the M-boxes, the L-boxes, and the combined detection catalogue. The evaluation is performed with respect to the full truth catalogue using only geometric bounding-box overlap (IoU = 0.5) and the \(C'_{\rm conf}\). The characterisation model predictions and the catalogue-scoring procedure are not used in the computation of these curves.}
    \label{fig:AP_50}
\end{figure}

Inference with the MSC characterisation model was performed at its optimal epoch (epoch 92) using a batch size of 256. For each prepared source, the model outputs five normalised parameters: $f_{\rm appr}$, $b_{\rm maj}$, $b_{\rm min}$, $\sin\phi$, and $\cos\phi$, all scaled to the range $[0,1]$. These were then rescaled to their physical units, and the position angle $\phi$ was recovered using
\[
\phi = \arctan\!\left(\frac{\sin\phi}{\cos\phi}\right).
\]
Since the MSC model was not trained to predict $C_{\rm frac}$, $C_{\rm spectral}$, and $C_{\rm profile}$, these parameters were fixed following \citet{Cornu2024} and Table~\ref{tab:combined_tables}, with values of 0.0375, 3, and 2, respectively. For consistency, we adopt the same notation as \citet{Cornu2024}. Results for the stand-alone SSD detector are presented in Section A, while the combined catalogue across all scales (SSD, MSD M-boxes, and MSD L-boxes) is presented in Section B of Table~\ref{tab:combined_results_equal_spacing}. The definitions of $F_{\rm sco}$, $N_{\mathrm{det}}$, $N_{\mathrm{match}}$, $N_{\mathrm{false}}$, and $\bar{w}$ are provided in Sect.~\ref{Scoring_procedure}. The quantity $N_{\mathrm{bad}}$ denotes the number of sources within $1.5 \times S'$ that show large positional offsets $D$, following \citet{Cornu2024}. Purity is calculated as
\begin{equation}
\mathrm{purity} = \frac{N_{\mathrm{match}}}{N_{\mathrm{det}}} \times 100\,\%.
\end{equation}

Section B presents the results of the merged catalogue corresponding to \(4 \times 4\) square box size and the contribution by the individual scale, such that \texttt{SSD} with $C'_{\rm conf}\geq0.5$ (top), \texttt{MSD} (M-boxes) with $C'_{\rm conf}\geq0.4$, and \texttt{MSD} (L-boxes) with $C'_{\rm conf}\geq0.3$, each evaluated at its optimal epoch. The final merged catalogue achieved a score of 483,875, with \(\bar{w}\) 0.8181 and a purity of 93.57\%. We found that the purity of the \texttt{SSD} decreased when the results of the \texttt{MSD} models were combined. This is because some sources detected by the \texttt{SSD} model were also identified by the \texttt{MSD} (M-boxes) model with higher \(C'_{\rm conf}\); during duplicate removal, the \texttt{MSD} detections replaced the \texttt{SSD} ones. However, false detections from both models do not decrease significantly and remain in the catalogue, reducing the purity. Despite this, the combined catalogue achieved a higher overall score because the $\bar{w}$ of matched detections outweighed the impact of false detections. The \texttt{MSD} (L-boxes) only marginally contributed to the final score because the large sources are very few in both the training set and the true catalogue. The scoring procedure also produces a matched catalogue containing all \(N_{\rm match}\) sources together with their corresponding source properties and parameter-wise score catalogue. For the best-performing catalogue (\(F_{\rm sco} = 483875\)), the values of $\bar{w}_{\rm position}$, $\bar{w}_{b_{\rm maj}}$, $\bar{w}_{b_{\rm min}}$, $\bar{w}_{f}$, $\bar{w}_{\phi}$, $\bar{w}_{c_{\rm spectral}}$, and $\bar{w}_{C_{\rm fraq}}$ are 0.976, 0.765, 0.717, 0.685, 0.624, 0.973, and 0.986, respectively.

Section~C of Table~\ref{tab:combined_results_equal_spacing} lists the results obtained by other teams for the 560\,MHz, 1000\,h dataset, together with our best configuration, following the notation of \citet{Cornu2024}. The proposed \texttt{YOLO-Chars} framework achieves performance comparable to the best previously published results on the SKA SDC1 benchmark, including MINERVA (YOLO-CIANNA). In particular, our best configuration attains a higher value of $\bar{w}$ and a slightly higher overall score, although differences remain in the individual detection metrics. These results indicate that \texttt{YOLO-Chars} can achieve competitive benchmark performance while adopting a different modelling strategy from previously published unified approaches.

In Section D, we show that increasing $C'_{\mathrm{conf}}$ in the combined catalogue reduces $F_{\mathrm{sco}}$, $N_{\mathrm{det}}$, $N_{\mathrm{match}}$, and $N_{\mathrm{false}}$, but improves both purity and $\bar{w}$. above the $C'_{\mathrm{conf}} \geq 0.6$ we observe that $F_{\mathrm{sco}}$ decreases to 412429 but purity increase to approximately 99\% and $\bar{w}$ to 0.8315. We observe a clear trade-off between source recovery and catalogue quality. Applying progressively higher \(C'_{\rm conf}\) to the final merged catalogue increases both purity and $\bar{w}$, but substantially reduces $N_{\mathrm{det}}$ and $N_{\mathrm{match}}$. We therefore adopt the original merged catalogue, constructed using $C'_{\mathrm{conf}}\geq0.5$ for S-boxes, $C'_{\mathrm{conf}}\geq0.4$ for M-boxes, and $C'_{\mathrm{conf}}\geq0.3$ for L-boxes, as our best configuration. Although applying a higher  $C'_{\mathrm{conf}}$ to the merged catalogue yields marginal improvements in purity and $\bar{w}$, it removes a significant number of valid detections, particularly from the MSD catalogues, resulting in a substantial reduction in $N_{\mathrm{match}}$ and the overall source recovery.

To further examine the source-characterisation performance of the MSC model, we analysed the physical properties $f$, $b_{\rm maj}$, $b_{\rm min}$, and $\phi$ using the matched catalogue corresponding to the best-performing configuration. Table~\ref{tab:characterisation_results} summarises the source-characterisation performance as a function of source SNR for both compact and extended sources, reporting the median fractional errors, $|\Delta f|/f_{\rm true}$, $|\Delta b_{\rm maj}|/b_{{\rm maj},\rm true}$, and $|\Delta b_{\rm min}|/b_{{\rm min},\rm true}$ (expressed as percentages), together with the mean absolute position-angle error, $\langle |\Delta\phi| \rangle$. The corresponding trends are shown in Fig.~\ref{fig:characterisation_vs_snr}. For all parameters, the errors generally decrease with increasing SNR. Extended sources exhibit larger errors than compact sources for flux density and source-size measurements, particularly at low SNR. In contrast, the $\phi$ errors are larger for compact sources, because in the catalogue most of the sources are below 1.5$\arcsec$, such that below the resolution limit or marginally resolved sources. We also evaluated the catalogue completeness as a function of true flux density for both the best-scoring catalogue (purity = 93.57 per cent) and a high-purity catalogue obtained using $C'_{\rm conf}\geq0.65$ (purity = 99.53 per cent). The completeness is defined as the fraction of true sources recovered within each true flux-density bin. The corresponding source-recovery and completeness curves are shown in Fig.~\ref{fig:completeness}. The best-scoring catalogue provides consistently higher completeness across most of the flux-density range, whereas the high-purity catalogue exhibits reduced completeness, particularly at lower flux densities, reflecting the expected trade-off between source recovery and catalogue purity when a more stringent confidence threshold is applied.

Additionally, we compute the average precision at IoU=0.5 ($AP_{50}$) using both the combined detection catalogue (SSD + MSD across M-boxes and L-boxes) and the individual catalogues from each scale on the full map from $C'_{\mathrm{conf}}$ from 0.05 to 1.0 for SSD and from 0.01 to 1.0 for M-boxes and L-boxes to evaluate performance across different thresholds. Detections below $C'_{\mathrm{conf}} = 0.05$ for SSD are discarded to reduce computational cost. The results are shown in Fig.~\ref{fig:AP_50}, which plots $AP_{50}$ for SSD (red), M-boxes (blue), L-boxes (green), and the combined catalogue (cyan). Predictions and truth catalogue exclude the training region to provide an unbiased estimate.

\section{Discussion} \label{sec:discussion}
The current MSC network outputs five intrinsic parameters: $f$, $b_{\rm maj}$, $b_{\rm min}$, $\sin(\phi)$, and $\cos(\phi)$, and could be extended to predict additional source properties such as $C_{\mathrm{frac}}$, $C_{\mathrm{spectral}}$, and $C_{\mathrm{profile}}$. The ability to accurately detect bboxes depends not only on the amount of training data but also on the choice of anchor boxes, which provide useful priors for bbox prediction. Overall, the results demonstrate that YOLO-style detectors can be effectively adapted to radio astronomical images, covering a wide range of angular scales and enabling a scalable pipeline for source detection. Compared with the detection networks, the characterisation network can remain relatively lightweight while still handling sources spanning a broad range of angular sizes. We also present several example $512 \times 512$ image fields from the full dataset in Appendix~\ref{appendix_2}, showing detections from the merged catalogue together with matched and false detections from the \texttt{SSD (S-boxes)}, \texttt{MSD (M-boxes)}, and \texttt{MSD (L-boxes)} models. Despite these encouraging results, several limitations remain.

\emph{\noindent\textbf{1. Training on idealised simulations:}} The detectors and the characterisation model were trained exclusively on idealised SKA SDC1 simulations, which lack real-world calibration artefacts, dynamic-range limitations, and residual PSF structures; therefore, generalisation to real observations is not guaranteed without further adaptation. The radio images from the surveys comprise a considerable portion of the astronomical data. Surveys from modern interferometers have grown in data size, including those from the SKA. More sensitive surveys exhibit numerous morphologically complex features; therefore, the treatment of extended sources requires special consideration. In our approach, training on specific ranges of angular size makes the network more versatile and scalable. Although trained solely on SDC1 simulations, the methodology is survey-agnostic and can also be adopted for radio surveys from existing radio telescopes. Applying it to real surveys (e.g.\ LOFAR [\citet{2017A&AShimwell}, \citet{2019A&AMahatma},\citet{Shimwell}, \citet{Gasperin}], MeerKAT [\citet{2024MNRASGoedhart}], ASKAP [\citet{2020PASAMcConnell}], VLA [\citet{1998AJCondon}], uGMRT [\citet{Intema}]) would require domain adaptation to account for differences in noise properties, dynamic variation in brightness, sensitivity, resolution, and source populations, and frequency dependence. Most radio sources follow power-law spectra $S_\nu \propto \nu^\alpha$ with $\alpha \lesssim -0.7$, causing their flux to drop rapidly with increasing frequency, while \(\sigma\) varies non-monotonically across bands. Our current approach will fail to provide better results for these spectral and noise variations.

\emph{\noindent\textbf{2. Characterisation of faint and complex sources:}} The MSC model is trained with apparent flux, which depends on the primary beam, and we also define the thresholds for \(f_{\rm appr}\), \(b_{\rm maj}\) and \(b_{\rm min}\) (see Section~\ref{Data_Description}), which biases characterisation performance against faint or barely resolved sources. Lowering the current apparent flux threshold ($1.6 \times 10^{-6}\ $Jy) would make the detector more sensitive to noise, leading to more false detections. Following this, the \(MSC\) model would also struggle to accurately characterise such faint sources. As a result, incorrect predictions and a reduced sub-average score will decrease the final score. Furthermore, the gt-bbox representation is inherently limited for multi-component or irregular sources such as FR\,I/II radio galaxies, for which segmentation-based or component-linking approaches may be required. In addition, the current framework does not account for variations in source size profiles (e.g., LAS, Gaussian versus exponential discs) or spectral classes (SS-AGNs, FS-AGNs, and SFGs). These aspects are left for future work.

{\emph{\noindent\textbf{3. Training-region selection:}} The choice of the region in the image for the training data is crucial and a limiting factor to better performance. As we use the training region defined in the official challenge, it does not represent the complete behaviour of the entire map. A more effective approach would be to sample training data uniformly throughout the map. Alternatively, it would be beneficial to simulate the data with multiple overlapping pointings, perform primary beam correction, and train the model more effectively to account for all variations on the map. This can improve detection performance, as the detector becomes sensitive to both clean and noisy regions. Since some sources may also be present in noisy areas, training in these regions can help reduce false detections and improve overall purity.

\emph{\noindent\textbf{4. Catalogue merging and source association:}} The catalogue-merging procedure is designed solely to remove duplicate detections arising from the overlap between the SSD and MSD catalogues and does not attempt to determine whether nearby radio components are physically associated. Consequently, closely spaced components belonging to the same radio source and unrelated neighbouring sources are treated identically during the merging process. Although the adopted $4\times4$ pixel square boxes were selected based on the sensitivity analysis presented in Table~\ref{tab:merge_box_size}, distinguishing between these scenarios requires additional morphology- and association-based analyses that are beyond the scope of this work. Future developments could incorporate more physically motivated source-association algorithms, such as centroid separation, source morphology, or probabilistic association, to improve the treatment of complex multi-component radio sources.

\emph{\noindent\textbf{5. Comparison with existing methods:}} The source-characterisation performance is evaluated directly against the ground-truth catalogue of the SKA SDC1 simulations, enabling an objective assessment of the recovery accuracy for flux density, source size, and position angle. Although a qualitative comparison between the \texttt{YOLO-CHARS} framework and PyBDSF is presented in Appendix~\ref{appendix_3} for a substantial subset of the SKA SDC1 field, a comprehensive quantitative benchmark against established source-finding and fitting pipelines over the entire field has not been undertaken. Such an analysis would require the construction of matched catalogues using consistent source-association, filtering, and evaluation procedures and is therefore left for future work.

In addition, the comparison with PyBDSF presented in Appendix~\ref{appendix_3} is restricted to treating PyBDSF as a complete source-finding and characterisation pipeline. Hybrid approaches, in which source detection is performed using \texttt{YOLO} and source characterisation is subsequently refined using traditional source-fitting techniques, have not been investigated in the present work. Assessing the potential benefits of such hybrid strategies would require dedicated source-association and fitting workflows and is therefore left for future study.

Furthermore, although the proposed framework adopts a decoupled architecture for source detection and source characterisation, the present work does not investigate how this design compares with unified end-to-end frameworks on challenging scientific use cases. Such a comparative study would provide further insight into the respective strengths and limitations of the two approaches and is therefore also left for future work. Future work will also explore alternative source-characterisation architectures, training strategies, and source-association methods to further improve parameter-recovery accuracy and catalogue quality.

\emph{\noindent\textbf{6. Future architectural developments:}} The framework adopts a YOLOv3-based architecture as a robust and well-established foundation for radio-source detection. More recent YOLO variants, including YOLOv4 \citep{bochkovskiy2020yolov4}, Ultralytics YOLOv5/v7/v8 \citep{yolov5, wang2022yolov7, yolov8_ultralytics}, and YOLOv9--v11 \citep{wang2024yolov9, THU_MIGyolov10, yolo11_ultralytics}, incorporate architectural developments such as anchor-free detection heads, enhanced backbone networks, attention mechanisms, and oriented bounding-box support. The potential benefits of these advances for radio-source detection have not been systematically evaluated in this work and warrant further investigation.

\section{Conclusion and summary} \label{sec:conclusion_and_summary}
In this work, we develop a custom YOLO-inspired detection framework for radio astronomical imaging and evaluate it on the simulated SKA SDC1 dataset at a single frequency. We adopted a disjoint multi-scale approach, separating networks for compact source detection and extended source detection. 
We demonstrate that the source detection problem can be viewed differently from the source characterisation problem. The main findings of this work are as follows.

\begin{enumerate}
    \item We have developed and evaluated a decoupled framework for source detection and source characterisation.
    \item We have implemented a multi-scale detection strategy designed to recover both compact and extended radio sources.
    \item We have developed a single multi-scale characterisation network capable of handling sources spanning a wide range of angular sizes.
    \item We have demonstrated competitive source-characterisation performance for \(f\), \(b_{\rm maj}\), \(b_{\rm min}\), and \(\phi\).
\end{enumerate}
More broadly, these findings highlight the importance of continued development of scale-adaptive, morphology-aware, and domain-robust source-finding frameworks for future SKA and SKA-pathfinder surveys.

A characteristic of the proposed framework is its modular design, which allows individual components to be trained, replaced, or extended independently without modifying the rest of the pipeline. Using the optimal configuration, \texttt{YOLO-CHARS} achieved an overall best score of 483\,875, corresponding to an average sub-score of \textbf{$\bar{w}=0.8181$} and a catalogue purity of 93.57\%. It also achieved the best purity score of 347\,236, corresponding to an average sub-score of \textbf{$\bar{w}=0.8388$} and a catalogue purity of 99.53\%. Corresponding to the best-score configuration, we also present the median fractional errors of \(f\), \(b_{\rm maj}\), and \(b_{\rm min}\), together with the mean absolute error in \(\phi\), across multiple SNR bins for both compact and extended sources, as well as the completeness comparison for catalogue purities of 93.57\% and 99.53\%. We also present the geometrical completeness and reliability corresponding to the optimal epochs using an IoU threshold of 0.5, for which the combined SSD and MSD detectors achieve an AP$_{50}$ of \textbf{15.279}

The automated source detection workflows can be tested using SKA data simulators \citep[e.g.][]{Dulwich2009, Sharma2026}. These simulators can produce observations with desired source populations, instrumental and imaging artefacts. Such simulation-based exercises also provide an effective framework for science verification. Our results demonstrate that the proposed framework provides a practical, scalable, and effective approach to automated source detection and characterisation in next-generation radio surveys like SKA, next-generation VLA, etc.

\section*{Acknowledgements}
The authors thank the anonymous referee for the careful review of the manuscript and for the constructive comments and suggestions, which have significantly improved the quality and clarity of this work. The authors also acknowledge the use of Google Colab computing resources during the development, training and evaluation of the models presented in this study. The authors thank Dr David Cornu for discussions on ML concepts.

\section*{Data Availability}

The data used in this work are publicly available\footnote{\url{https://www.skao.int/en/464/ska-science-data-challenge-1}}. The catalogues generated and analysed in this study are publicly available on Zenodo (\url{https://doi.org/10.5281/zenodo.18081815}).
\bibliographystyle{mnras}
\bibliography{example}

\appendix
\appendix

\section{Epoch Analysis} \label{epoch_analysis}

\begin{figure}
    \centering
    \includegraphics[width=1.0\linewidth]{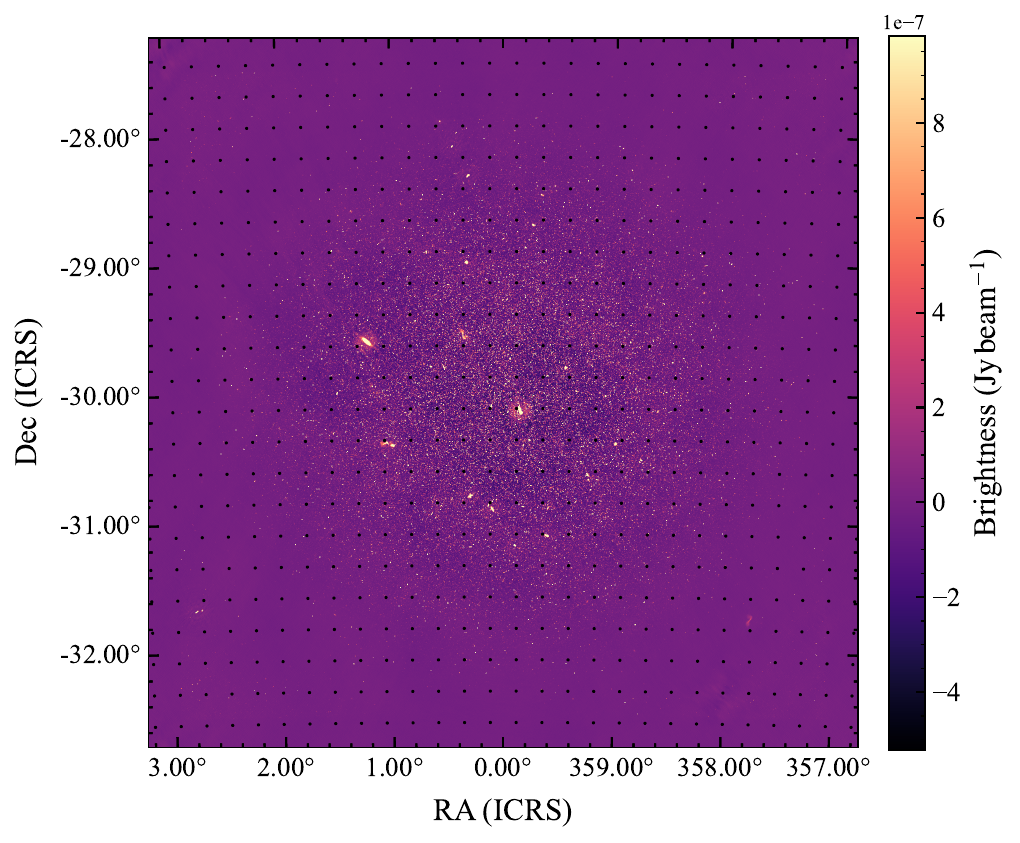}
    \caption{
        728 Sampling grid points on a $32768 \times 32768$ full image. Each grid point represents the RA and Dec centre of a $512 \times 512$ image cutout in FITS format, which preserves the coordinate details, transformed to pixel space via WCS. These cutouts are used to determine the optimal generalised epoch for all three models.
    }
    \label{fig:patch_overlay_grid}
\end{figure}

\begin{figure}
    \centering
    \includegraphics[width=1.\linewidth]{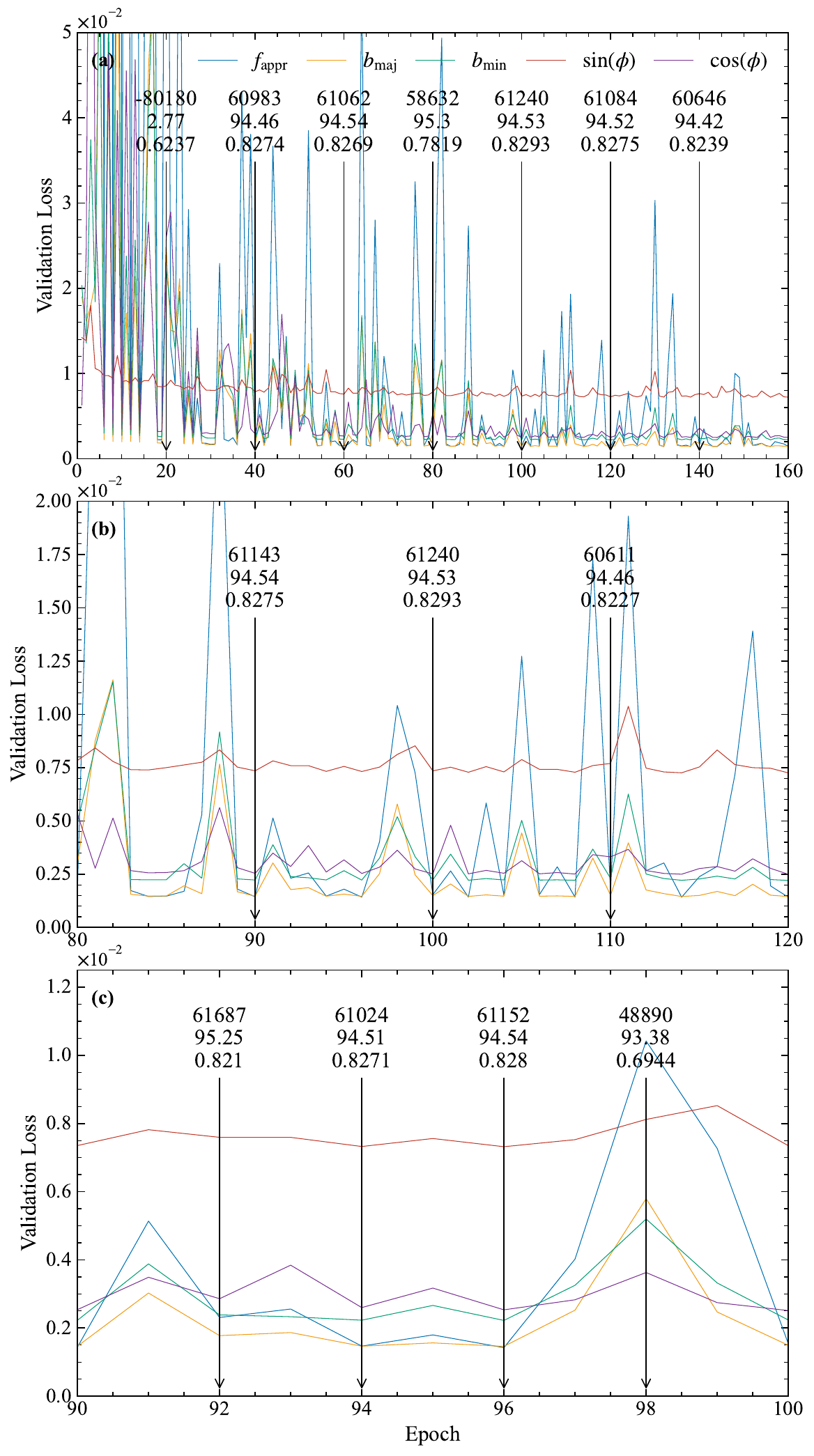}
    \caption{Score optimisation across MSC epochs with validation loss curves for the five predicted parameters: $f_{\rm appr}$, $b_{\rm maj}$, $b_{\rm min}$, $\sin(\phi)$, and $\cos(\phi)$. The SSD model is fixed at its final epoch (40). Arrows at selected epochs indicate the MSC model's performance metrics: total score (top label), purity (middle label), and average accuracy (bottom label). The top panel shows optimisation at 20-epoch intervals (epochs 20--140), the middle panel at 10-epoch intervals (epochs 90--110), and the bottom panel at 2-epoch intervals (epochs 92--98). The maximum score is achieved at epoch 92 (bottom panel).}
    \label{fig:val_loss_annotated}
\end{figure}

\begin{figure}
    \centering
    \includegraphics[width=1.\linewidth]{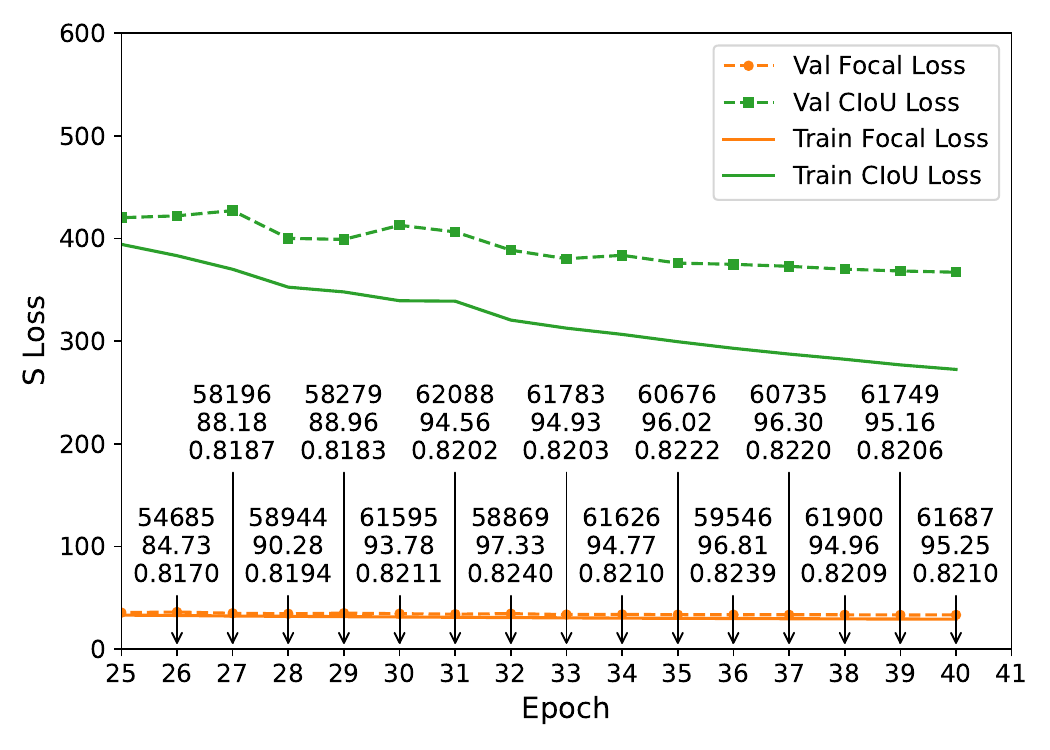}
    \caption{Score optimisation across SSD epochs with the MSC model weights fixed at epoch 92. The first 25 epochs are skipped, and scores accuracy and purity are evaluated for epochs 26--40, showing validation and training CIoU and Focal loss, respectively. Epoch 31 achieves the maximum \(F_{\rm sco}\), while epoch 38 shows the second-highest \(F_{\rm sco}\) with higher purity and \(\bar{w}\) compared to epoch 31.}
    \label{fig:val_loss_detection_components_det_S}
\end{figure}

\begin{figure}
    \centering
    \includegraphics[width=1.0\linewidth]{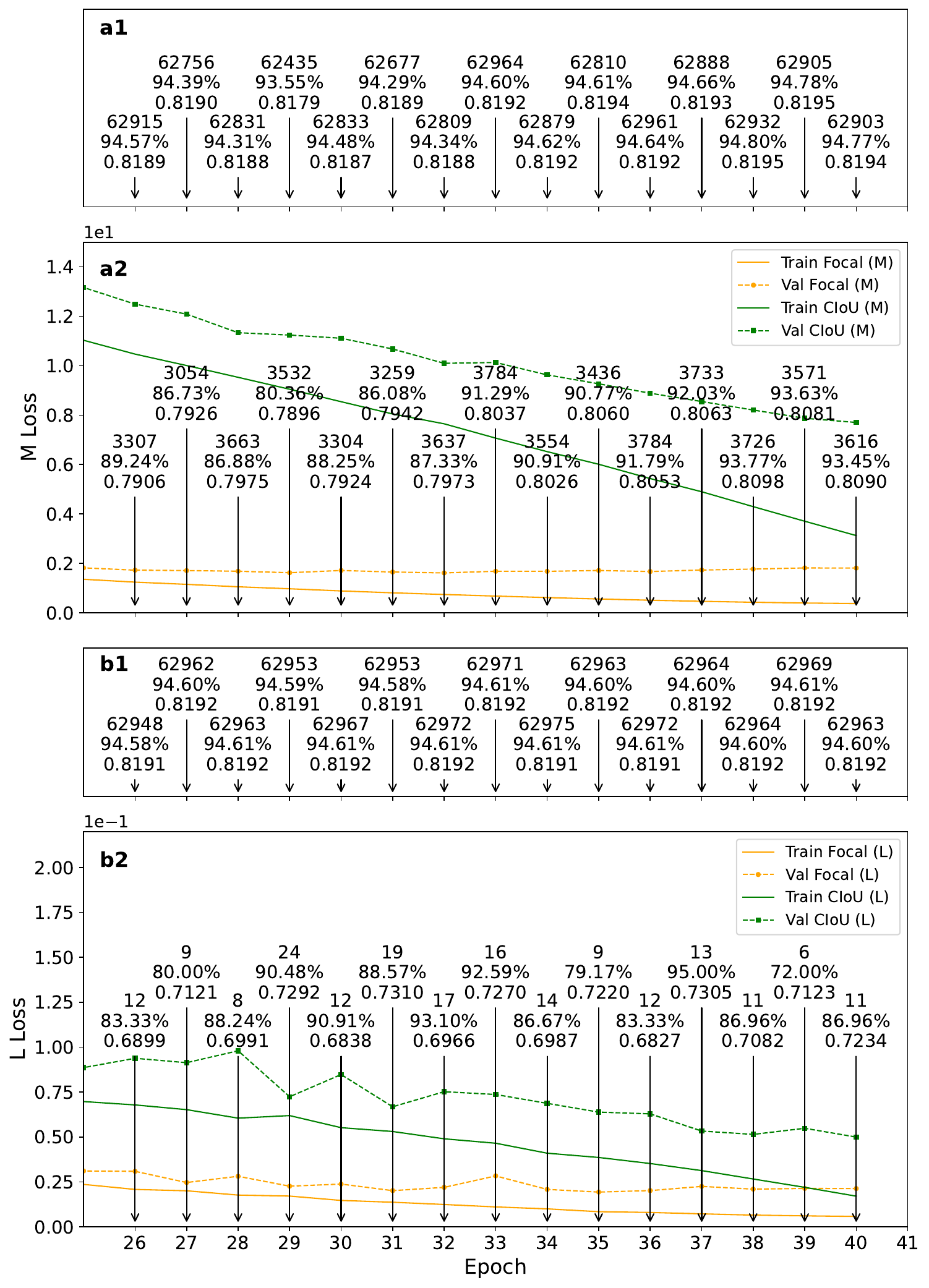}
    \caption{Combined score, purity, and accuracy across MSD epochs 26--40. Panel (a1) shows M-boxes results, including fixed L-boxes results at epoch~40, while panel (b1) shows L-boxes results, including fixed M-boxes results at epoch~33. Both include SSD (S-boxes) results corresponding to epoch~38. Panels (a2) and (b2) display individual M-boxes and L-boxes results, respectively. The highest combined score of 62,975 was achieved at epochs 33, 34, 38, and 92 for M-boxes, L-boxes, S-boxes, and MSC, respectively, as shown in panel (b1) at epoch 34.}
    \label{fig:val_loss_detection_components_det_M_L}
\end{figure}
Following training, the optimal epochs for the \texttt{SSD}, \texttt{MSD} (M-boxes and L-boxes) and \texttt{MSC} models were identified using a procedure similar to that described in Sect.~\ref{Analysis_and_Results}. The key difference is that inference was performed at multiple epochs for each model on a reduced subset of the full map to limit computational cost. This subset spans RA from $-3.2^\circ$ to $3.2^\circ$ (28 points) and Dec from $-32.8^\circ$ to $-27.2^\circ$ (24 points), yielding 672 positions in total (see Fig.~\ref{fig:patch_overlay_grid}). The procedure is outlined below.

\noindent\textbf{Step 1:} First, the optimal epoch was determined for the \texttt{MSC} model. For this purpose, the \texttt{SSD} model was fixed at its final epoch (epoch~40), which was assumed to provide stable source detections. Source detection was performed at this epoch, while source characterisation with the \texttt{MSC} model was evaluated at intervals of 20 epochs, as early epochs exhibited suboptimal performance. At each epoch, a catalogue was generated following the procedure described in Sect.~\ref{Analysis_and_Results} and assessed using the scoring method. The evolution of the metrics $F_{\rm{sco}}$, purity and $\bar{w}$ was monitored (top panel of Fig.~\ref{fig:val_loss_annotated}). We found that $F_{\rm{sco}}$ increased and then decreased between epochs 80 and 120. A finer search over this range, evaluated at every epoch, identified epoch~92 as the optimal configuration, yielding $ F_{\rm{sco}} = 61\,687$, a purity of 95.25\% and $\bar{w} = 0.821$, in combination with the \texttt{SSD} model at epoch~40.

\noindent\textbf{Step 2:} The \texttt{MSC} model was then fixed at epoch~92, and inference with the \texttt{SSD} model was performed at each epoch starting from epoch~25, as previous epochs produced suboptimal results. The metrics $F_{\rm sco}$, purity and $\bar{w}$ were evaluated at each epoch (see Figure~\ref{fig:val_loss_detection_components_det_S}). We found that the \texttt{SSD} model achieved its best performance at epoch~38, with $F_{\rm sco} = 61900$, a purity of 94.96\% and $\bar{w} = 0.8209$. This configuration was therefore adopted as the optimal combination of the \texttt{SSD} and \texttt{MSC} models.

\noindent\textbf{Step 3:} To determine the optimal generalisation epochs for the M-boxes and L-boxes in the \texttt{MSD} model, the \texttt{SSD} and \texttt{MSC} models were fixed at epochs~38 and~92, respectively. The L-boxes were first fixed at epoch~40, while the M-boxes were evaluated throughout epochs~25--39, with \texttt{SSD} detections at epoch~38 included in the merged catalogue. The merged catalogue reached a maximum $F_{\rm{sco}}$ at epoch~33, which was therefore selected as the optimal epoch for the M-boxes (see panels~a1 and~a2 in Fig.~\ref{fig:val_loss_detection_components_det_M_L} for the combined and individual contributions, respectively). The same procedure was then applied to the L-boxes by varying their epochs from~25 to~39 while keeping the M-boxes fixed at epoch~33. The corresponding combined and individual results are shown in panels~b1 and~b2 of Fig.~\ref{fig:val_loss_detection_components_det_M_L}, respectively.

\section{Example Images} \label{appendix_2}
\begin{figure*}
    \centering
    \includegraphics[width=1.\linewidth]{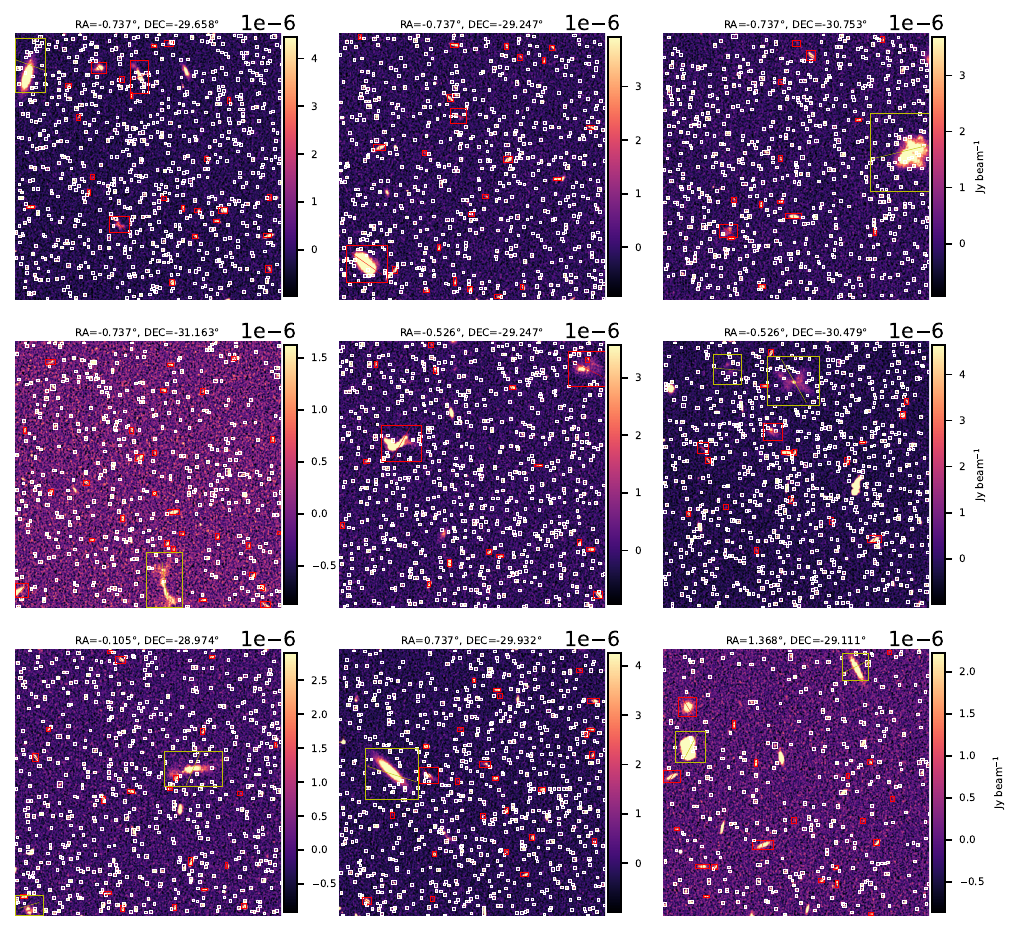}
    \caption{
    Example of 9 fields showing matched sources identified using the SKA SDC1 scoring procedure, corresponding to their $\phi$. The \texttt{SSD} (S-boxes; white), \texttt{MSD} (M-boxes; red), and \texttt{MSD} (L-boxes; yellow) models. Image-centre coordinates are annotated in RA and Dec.}
    \label{fig:match_dets}
\end{figure*}

\begin{figure*}
    \centering
    \includegraphics[width=1.\linewidth]{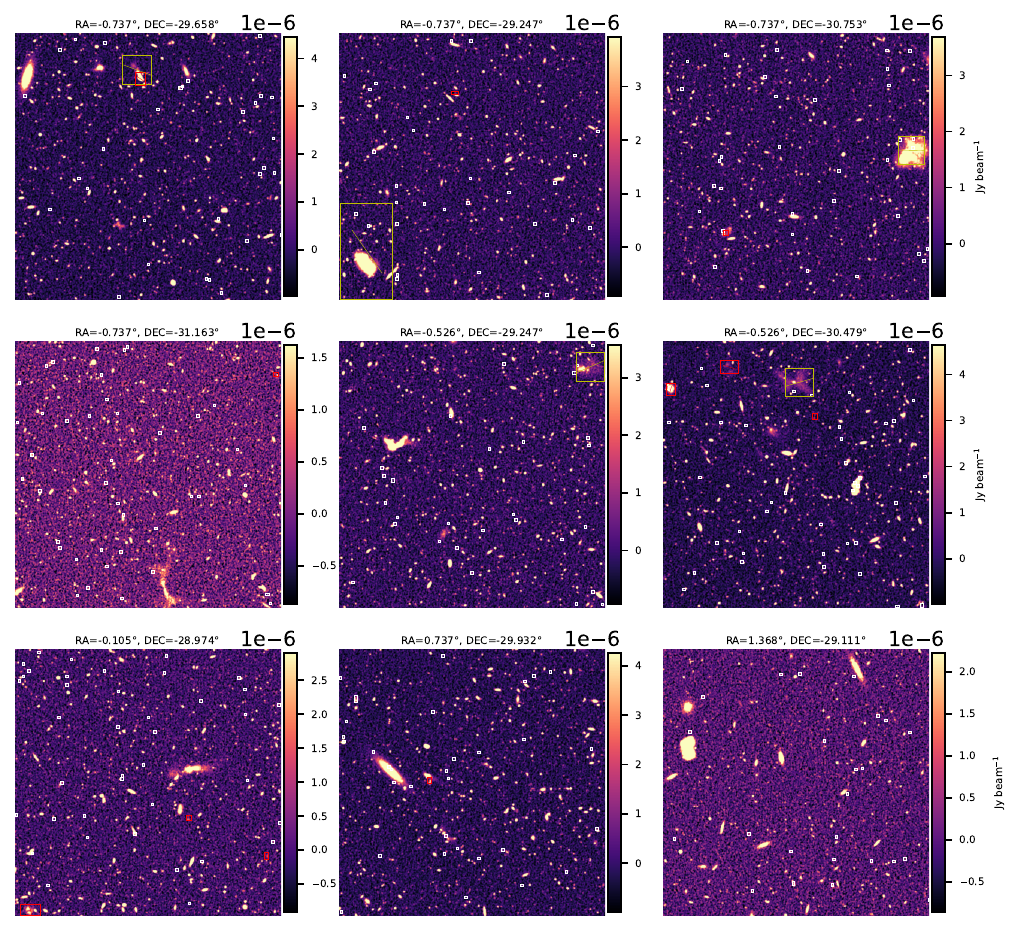}
    \caption{Example of the same 9 fields as in  Fig.~\ref{fig:match_dets} showing bad/false detections according to the scoring procedure with corresponding to their $\phi$. Predicted bbox colours follow the same convention as in Fig.~\ref{fig:match_dets}.}
    \label{fig:unmatch_dets}
\end{figure*}
This appendix presents several example $512 \times 512$ image patches extracted from the full field and normalised using Equation~\eqref{eq:Det_Normalization}. Each image shows detected sources with predicted gt-bboxes overlaid from our models.  
The results include detections from the SSD, MSD (M-boxes) and MSD (L-boxes) models, along with matched detections (see Figure~\ref{fig:match_dets}) and false or bad detections (see Figure~\ref{fig:unmatch_dets}), evaluated using the SKA SDC1 scoring procedure.  
In the visualisations, white boxes indicate SSD detections, while magenta and yellow boxes represent the MSD M-boxes and L-boxes outputs, respectively. A confidence threshold of 0.5 is used for SSD detections and 0.4 and 0.3 for MSD M-boxes and L-boxes detections, respectively.

\section{COMPARISON OF PyBDSF AND YOLO-CHARS} \label{appendix_3}

In this appendix, we compare the proposed decoupled \texttt{YOLO-CHARS} framework with PyBDSF, a widely used radio source-finding and characterisation package. The analysis is performed on the central $2.738^\circ \times 2.35^\circ$ region of the SKA SDC1 field (Fig.~\ref{fig:unified_testing_field}) and is intended to provide a representative benchmark against a classical source-finding approach. The truth catalogue is restricted to this evaluation region, with sources located outside the region and within the training area removed to ensure that all performance metrics, completeness measurements, and characterisation analyses are computed using an independent test sample.

PyBDSF was executed using a hard detection threshold with \texttt{thresh\_pix}=5.0 and \texttt{thresh\_isl}=4.0. The local RMS map
was estimated using \texttt{rms\_box}=(75,20) pixels with adaptive RMS estimation disabled (\texttt{adaptive\_rms\_box=False}). A fixed circular restoring beam corresponding to the SDC1 simulation ($b_{\rm maj}=b_{\rm min}=1.5''$, $\phi=0^\circ$) was adopted, while PSF variation (\texttt{psf\_vary\_do=False}) and wavelet decomposition (\texttt{atrous\_do=False}) were disabled. The PyBDSF catalogue was generated from the primary-beam-corrected image, whereas \texttt{YOLO-CHARS} evaluated using the non-primary-beam-corrected image. 

The results presented in this appendix show that \texttt{YOLO-CHARS} achieves performance comparable to PyBDSF over the common evaluation region, while attaining higher performance for several evaluation metrics. For the benchmark configuration adopted in Table~\ref{tab:best_model_epoch_for_testing_region}, \texttt{YOLO-CHARS} achieves a catalogue purity comparable to that of PyBDSF while recovering a larger number of matched sources and yielding a higher value of $\bar{w}$. Figure~\ref{fig:catalog_comparison_histograms} compares the $\Delta f/f_{\rm true}$, $\Delta b_{\rm maj}/b_{\rm maj, true}$, $\Delta b_{\rm min}/b_{\rm min,true}$, and $\Delta\phi$
for \texttt{YOLO-CHARS} and PyBDSF. While PyBDSF achieves a higher $\bar{w}_{f}$, \texttt{YOLO-CHARS} attains higher $\bar{w}_{b_{\rm maj}}$, $\bar{w}_{b_{\rm min}}$, and $\bar{w}_{\phi}$, indicating more accurate estimation of the source size and orientation parameters over the common evaluation region. Although \texttt{YOLO-CHARS} recovers a larger number of faint matched sources than PyBDSF, the completeness curves (Fig.~\ref{fig:flux_completeness_comparison_uni_py_dec}) show that PyBDSF achieves higher completeness over parts of the higher-SNR regime. This behaviour suggests that \texttt{YOLO-CHARS} is more effective at recovering faint sources, whereas PyBDSF performs better for a subset of brighter sources. Further improvements in flux-density estimation may therefore provide an effective route to enhancing the overall performance of the proposed framework. A comprehensive comparison with optimised PyBDSF configurations and other classical source-finding pipelines is left for future work.

\begin{figure}
    \centering
    \includegraphics[width=\linewidth]{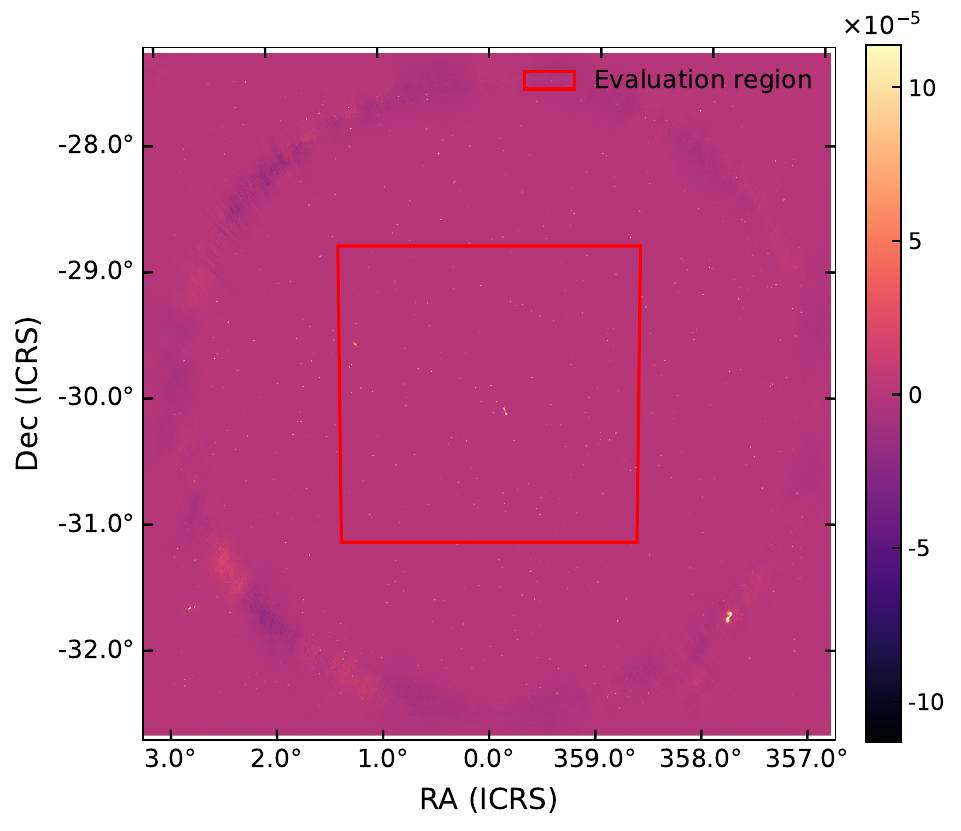}
    \caption{Primary-beam-corrected SKA SDC1 image showing the evaluation region (red rectangle outlines) used for the benchmark comparison between \texttt{YOLO-CHARS} and PyBDSF. The selected region spans $2.738^\circ$ in RA and $2.35^\circ$ in Dec.}
    \label{fig:unified_testing_field}
\end{figure}

\begin{table}
\centering
\scriptsize
\setlength{\tabcolsep}{4pt}
\caption{Performance metrics for the benchmark comparison between the decoupled \texttt{YOLO-CHARS} framework and PyBDSF on the common evaluation regime shown in Fig.~\ref{fig:unified_testing_field}. For the decoupled framework, a detection confidence threshold of $C'_{\rm conf}\geq0.64$ is adopted, resulting in a catalogue purity of 99.43\%. The subsequent completeness and characterisation analyses presented in this appendix are based on these configurations.}
\begin{tabular}{cccccccc}
\hline
Model($C_{\rm{conf}}$) & $F_{\rm{sco}}$ & $N_{\rm det}$ & $N_{\rm match}$ & $N_{\rm false}$ &
$N_{\rm bad}$ & Purity (\%) & $\bar{w}$  \\
\hline
PyBDSF(--) & 194996 & 251902 & 250382 & 1520 & 922 &  99.39 & 0.7848\\
\textbf{YOLO-CHARS (0.64)} & \textbf{282220} & \textbf{341125} & \textbf{339192} & \textbf{1933} & \textbf{404} & \textbf{99.43} & \textbf{0.8377}\\
\hline
\end{tabular}
\label{tab:best_model_epoch_for_testing_region}
\end{table}

\begin{figure}
    \centering
    \includegraphics[width=\linewidth]{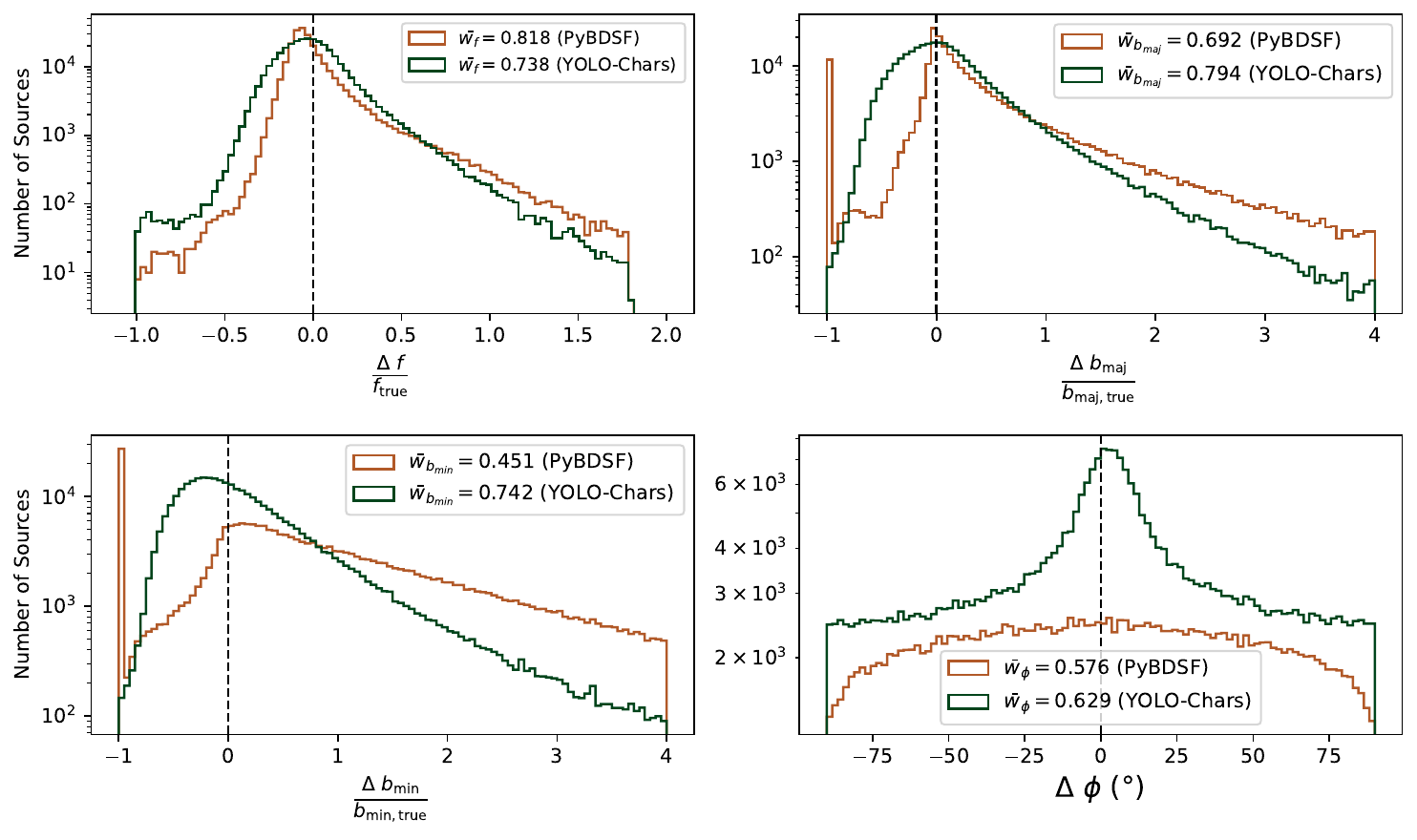}
    \caption{Comparison of the source-characterisation residual distributions obtained with the \texttt{YOLO-CHARS} framework and PyBDSF for the benchmark configurations listed in Table~\ref{tab:best_model_epoch_for_testing_region}. The panels show the fractional flux-density residual, $\Delta f/f_{\rm true}$, the fractional major-axis residual, $\Delta b_{\rm maj}/b_{\rm maj, true}$, the fractional minor-axis residual, $\Delta b_{\rm min}/b_{{\rm min},\rm true}$, and the position-angle residual, $\Delta\phi$. The y-axis is displayed on a logarithmic scale.}
    \label{fig:catalog_comparison_histograms}
    
\end{figure}

\begin{figure}
    \centering
    \includegraphics[width=\linewidth]{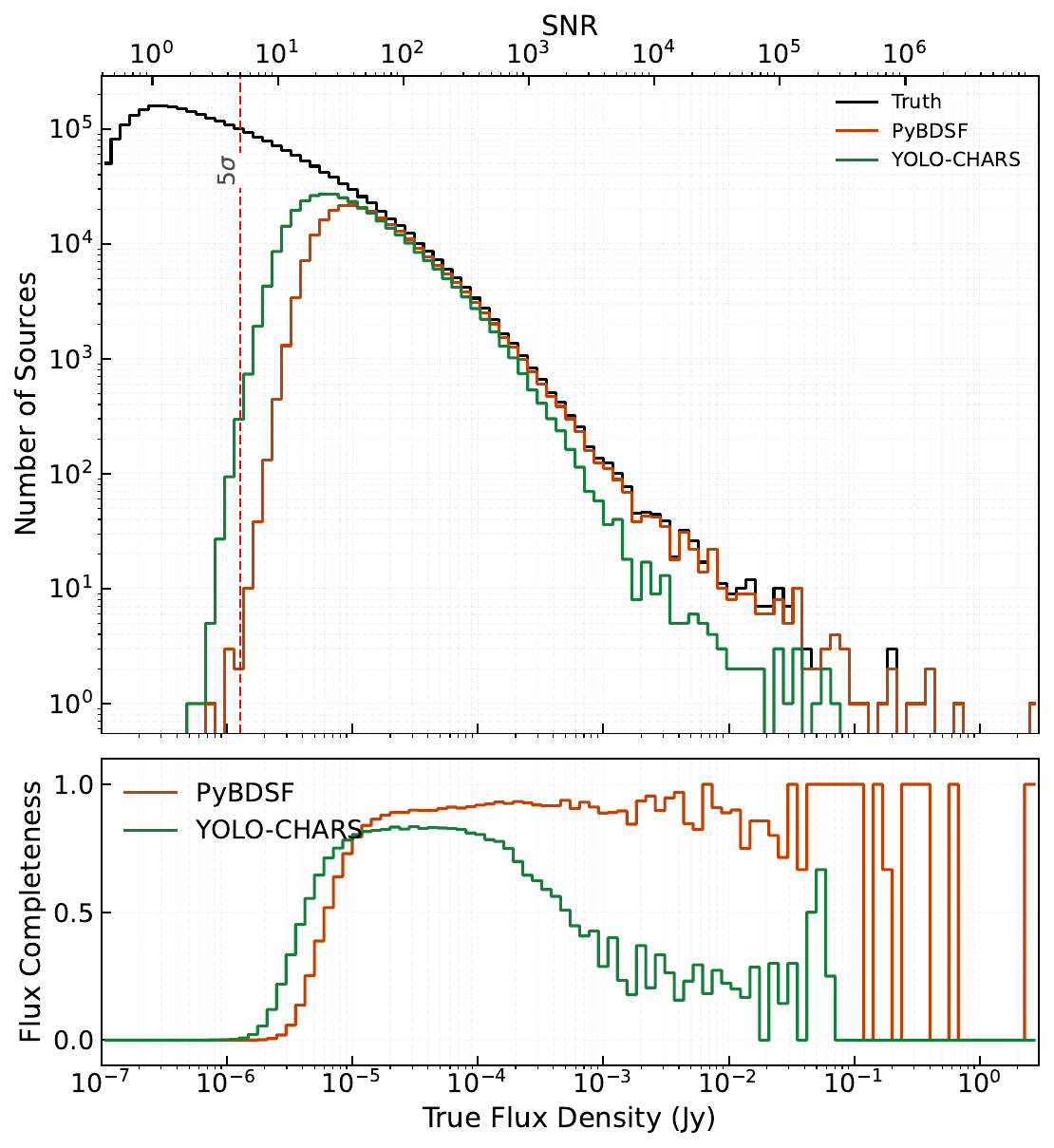}
    \caption{Completeness comparison for the \texttt{YOLO-CHARS} framework and PyBDSF using the benchmark configurations adopted in Table~\ref{tab:best_model_epoch_for_testing_region}. The upper panel shows the number of sources in each true-SNR bin for the truth catalogue and the corresponding number of matched sources recovered by each method. The lower panel shows the completeness, defined as the fraction of truth-catalogue sources recovered within each true-SNR bin, i.e. $N_{\rm match}/N_{\rm true}$.}
    \label{fig:flux_completeness_comparison_uni_py_dec}
\end{figure}
\label{lastpage}
\end{document}